\documentclass[journal=ancac3,manuscript=article]{achemso}

\usepackage{hyperref}
\hypersetup{
    colorlinks=true,
    linkcolor=blue,
    citecolor=blue,      
    urlcolor=blue,
    pdfpagemode=FullScreen,
    }
\usepackage{graphicx}
\usepackage{amsmath}
\usepackage{wasysym}
\usepackage{amsfonts}
\usepackage{color}
\usepackage{verbatim}
\usepackage{upgreek}
\usepackage{bm}
\usepackage{textcomp}
\usepackage[symbol*]{footmisc}
\usepackage{float}
\usepackage{braket}
\usepackage{soul}
\usepackage{dsfont}
\newcommand{\bvec}[1]{\boldsymbol{#1}} 

\DefineFNsymbolsTM{otherfnsymbols}{
  \textdagger    \dagger
  \textdaggerdbl \ddagger
  \textsection   \mathsection
  \textparagraph \mathparagraph
}
\setfnsymbol{otherfnsymbols}

\author{Konstantin G.~Wirth} 
\email{konstantin.wirth@rwth-aachen.de}
\affiliation{$1^{st}$ Institute of Physics, RWTH Aachen University, 52074 Aachen, Germany}
\author{Jonas B.~Hauck} 
\affiliation{Institute for Theory of Statistical Physics, RWTH Aachen University, and JARA Fundamentals of Future Information Technology, 52062 Aachen, Germany}
\author{Alexander Rothstein} 
\affiliation{$2^{nd}$ Institute of Physics and JARA-FIT, RWTH Aachen University, 52074 Aachen, Germany}
\alsoaffiliation{Peter Gr\"unberg Institute (PGI-9), Forschungszentrum J\"ulich, 52425 J\"ulich, Germany}
\author{Hristiyana Kyoseva}
\affiliation{$2^{nd}$ Institute of Physics and JARA-FIT, RWTH Aachen University, 52074 Aachen, Germany}
\author{Dario Siebenkotten}
\affiliation{$1^{st}$ Institute of Physics, RWTH Aachen University, 52074 Aachen, Germany}
\author{Lukas Conrads} 
\affiliation{$1^{st}$ Institute of Physics, RWTH Aachen University, 52074 Aachen, Germany}
\author{Lennart Klebl} 
\affiliation{Institute for Theory of Statistical Physics, RWTH Aachen University, and JARA Fundamentals of Future Information Technology, 52062 Aachen, Germany}
\author{Ammon Fischer} 
\affiliation{Institute for Theory of Statistical Physics, RWTH Aachen University, and JARA Fundamentals of Future Information Technology, 52062 Aachen, Germany}
\author{Bernd Beschoten}
\affiliation{$2^{nd}$ Institute of Physics and JARA-FIT, RWTH Aachen University, 52074 Aachen, Germany}
\author{Christoph Stampfer}
\affiliation{$2^{nd}$ Institute of Physics and JARA-FIT, RWTH Aachen University, 52074 Aachen, Germany}
\alsoaffiliation{Peter Gr\"unberg Institute (PGI-9), Forschungszentrum J\"ulich, 52425 J\"ulich, Germany}
\author{Dante M. Kennes}
\affiliation{Institute for Theory of Statistical Physics, RWTH Aachen University, and JARA Fundamentals of Future Information Technology, 52062 Aachen, Germany}
\alsoaffiliation{Max Planck Institute for the Structure and Dynamics of Matter, Center for Free Electron Laser Science, Hamburg, Germany}
\author{Lutz Waldecker} 
\affiliation{$2^{nd}$ Institute of Physics and JARA-FIT, RWTH Aachen University, 52074 Aachen, Germany}
\author{Thomas Taubner}
\email{taubner@physik.rwth-aachen.de}
\affiliation{$1^{st}$ Institute of Physics, RWTH Aachen University, 52074 Aachen, Germany}

\title[]{
Experimental observation of ABCB stacked tetralayer graphene
}

\begin{document}

\begin{abstract}

In tetralayer graphene, three inequivalent layer stackings should exist, however, only rhombohedral (ABCA) and Bernal (ABAB) stacking have so far been observed. 
The three stacking sequences differ in their electronic structure, with the elusive third stacking (ABCB) being unique as it is predicted to exhibit an intrinsic bandgap as well as locally flat bands around the K points.
Here, we use scattering-type scanning near-field optical microscopy and confocal Raman microscopy to identify and characterize domains of ABCB stacked tetralayer graphene.
We differentiate between the three stacking sequences by addressing characteristic interband contributions in the optical conductivity between 0.28 and 0.56~{eV} with amplitude and phase-resolved near-field nano-spectroscopy.
By normalizing adjacent flakes to each other, we achieve good agreement between theory and experiment, allowing for the unambiguous assignment of ABCB domains in tetralayer graphene.
These results establish near-field spectroscopy at the interband transitions as a semi-quantitative tool, enabling the recognition of ABCB domains in tetralyer graphene flakes and therefore, providing a basis to study correlation physics of this exciting new phase.

\end{abstract}

\begin{figure}[!bth]
    \begin{center}
        \includegraphics[width=0.75\textwidth]{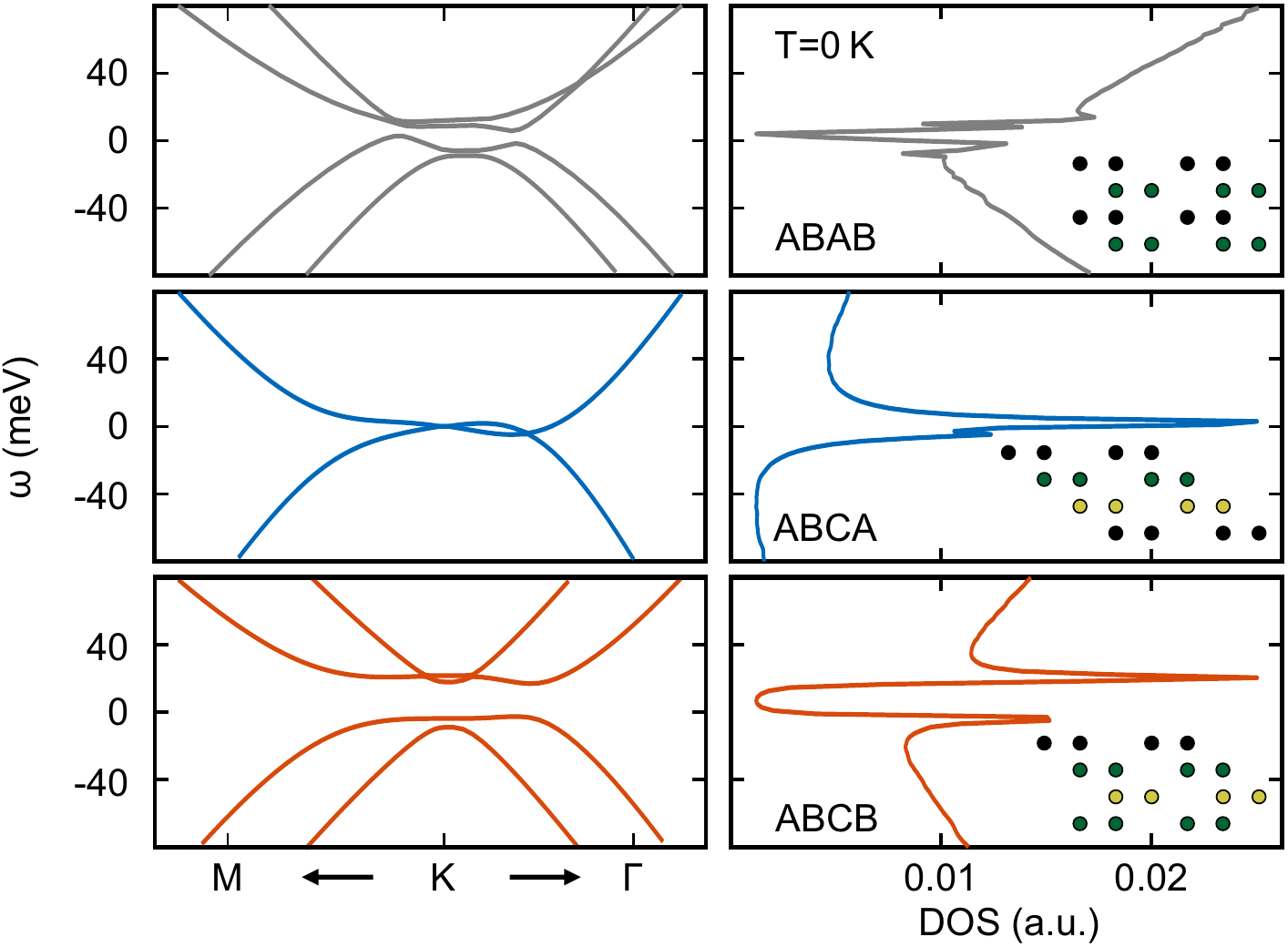}
        \caption{\textbf{Low-energy electronic properties of the three 4LG stackings.} Low-energy-band structures and density of states (DOS) of tetralayer graphene for three different stacking orders without an applied electric field from the tight-binding model described in the text. Upper panel ABAB, central panel ABCA and lower panel ABCB. The small insets show the side view of the atomic stackings. Note that the band touching point in ABAB graphene is not correctly captured by this simple SWMC-model (see S4). A very small band gap is shown in the upper panel, which does not exist in reality. We introduced an artificial broadening of $1$\;meV in the calculation of the DOS. 
        }
        \label{fig:0}
    \end{center}
\end{figure}

The crystallographic stacking order of few layer graphene (FLG) greatly influences its electronic and optical properties. 
Naturally occurring crystallographic structures of graphene host interesting phenomena such as quantum Hall states in single and bilayer graphene~\cite{Zhang2005Nov, Novoselov2006Mar}, with Bernal stacked bilayer graphene exhibiting superconductivity upon applying a magnetic field~\cite{zhou_isospin_2021}. 
Half- and quarter-metals have been reported for rhombohedral stacked trilayer graphene~\cite{zhou_superconductivity_2021, zhou_half-_2021}, while a charge-transfer excitonic insulator and a ferrimagnet are candidates for phases of matter realized in rhombohedral four-layer (ABCA) graphene~\cite{kerelsky_moireless_2021}.
In addition, artificial stackings of graphene layers with a twist angle have been shown to lead to flat bands and correlated phenomena~\cite{Balents2020,kennes_moire_2021} such as unconventional superconductivity in twisted bilayer graphene~\cite{cao_unconventional_2018} or ferromagnetic insulating states in twisted double bilayer graphene~\cite{liu_tunable_2020, shen_correlated_2020}. 
The most common stacking in FLG is Bernal stacking, which is the energetically favorable configuration~\cite{aoki_dependence_2007}, while rhombohedral stacking is less common~\cite{zhang_light-induced_2020,lui_imaging_2011}.
For tetralayer graphene (4LG), besides rhombohedral and Bernal (ABAB) stackings, also a mixed stacking has been predicted to be metastable, namely the equivalent stackings ABCB and ABAC (here denoted as ABCB)~\cite{aoki_dependence_2007, latil_charge_2006}, c.f. insets Figure~\ref{fig:0}.

ABCB stacked 4LG exhibits a unique electronic band structure~\cite{aoki_dependence_2007, latil_charge_2006, min_electronic_2008}:
It is the thinnest graphene-based intrinsic band insulator. 
It is expected to have a band gap of 8.8~{meV}~\cite{latil_charge_2006} which, upon application of an out-of-plane electric displacement field, should close~\cite{aoki_dependence_2007}. 
In addition, it is predicted to feature strong van-Hove singularities at the band edges.
This is shown in Figure~\ref{fig:0}, which displays the low-energy bandstructure, density of states (DOS) and a side view of the three possible crystallographic stackings of 4LG: ABAB, ABCA and ABCB.
The electronic structure of ABCB is neither related to the one of Bernal nor to the one of rhombohedral stacked graphene and its low energy spectrum consists of a locally flat (approximately cubic) band intersected by a massive Dirac cone~\cite{min_electronic_2008}, promoting the strong van-Hove singularities found.
The flat bands at the valleys $K$ and $K^{\prime}$, which emerge in ABCB graphene, might render this stacking particularly intriguing from the viewpoint of correlated phenomena.
This stacking, however, has so far eluded experimental observation.

Here, we provide experimental evidence for ABCB stacked tetralayer graphene by amplitude and phase resolved infrared nano-spectroscopy with scattering-type scanning near-field optical microscopy (s-SNOM), by addressing characteristic interband contributions in the optical conductivity.
Normalizing the amplitude and phase values to the respective values of ABAB stacking, which is relatively featureless in the investigated energy range, overcomes the issue of the influence of the unknown tip shape on our spectra. We achieve quantitative agreement between measured and calculated amplitude and phase spectra. 
Furthermore, we provide measurements of the Raman G, 2D and M peaks of all three stacking orders, which constitutes a second, independent means of unambiguous identification of the different domains.
This remedies the long-standing elusive nature of the ABCB stacking order and will provide a basis for studying the extraordinary electronic properties of this stacking.

\section{Results}

\subsection{Domain imaging}
To characterize different graphene stacking orders, optical techniques offer a wide range of tools. 
Absorption measurements in the infrared regime between $0.2$ and $0.9$~{eV} of FLG give access to characteristic absorption peaks around the interband transitions, which can be directly linked to the electronic structure~\cite{mak_evolution_2010}, and have thus been used to distinguish ABCA and ABAB~\cite{mak_electronic_2010} stacked 4LG. 
Confocal Raman spectroscopy is another standard tool for identifying stacking orders in FLG~\cite{lui_observation_2011}. 
More advanced spectroscopy methods, such as Magneto-Raman, offer even better characterisation capabilities for the stacking order, as they are sensitive to electronic bands near the Fermi edge \cite{berciaud_probing_2014, henni_rhombohedral_2016}.
All of these methods, however, are limited in spatial resolution by diffraction.
Thus, common far-field spectroscopy methods are prone to overlook small domains on these flakes, in particular those of unknown crystallographic stackings.

The diffraction limit can be overcome by employing infrared scattering type scanning near-field optical microscopy (s-SNOM), which enables infrared nano-imaging with a lateral resolution down to 20~{nm}~\cite{taubner_performance_2003}.
s-SNOM enabled the real-space observation of surface plasmon polaritons (SPPs) in graphene~\cite{fei_gate-tuning_2012, chen_optical_2012}. 
With the help of SPP reflection at inhomogeneities, s-SNOM has also been used to image grain boundaries~\cite{fei_electronic_2013}, and solitons in FLG~\cite{jiang_soliton-dependent_2016, jiang_plasmon_2017, jiang_manipulation_2018} as well as twisted bilayer graphene~\cite{sunku_photonic_2018} and other moir\'e heterostructures~\cite{halbertal-2021-metrology}.
At photon energies above 0.2 eV, s-SNOM gives access to the stacking-specific optical conductivities of FLG, and the scattering amplitude and phase values can be retrieved simultaneously. 
This allows to assign  sub-diffraction FLG domains to stackings by recording images at specific photon energies~\cite{kim_stacking_2015, jeong_mapping_2017}. 
Recently, it has been shown for IR nano-spectroscopy on bilayer graphene around the conductivity resonances at $0.39$~{eV} that optical amplitude and phase scale with the characteristics of real and imaginary part of the conductivity~\cite{wirth_tunable_2021}. 
However, the quantitative agreement between theoretical prediction and experimental data in these studies is still lacking.

\begin{figure*}[!bth]
    \begin{center}
        \includegraphics[width=1.0\textwidth]{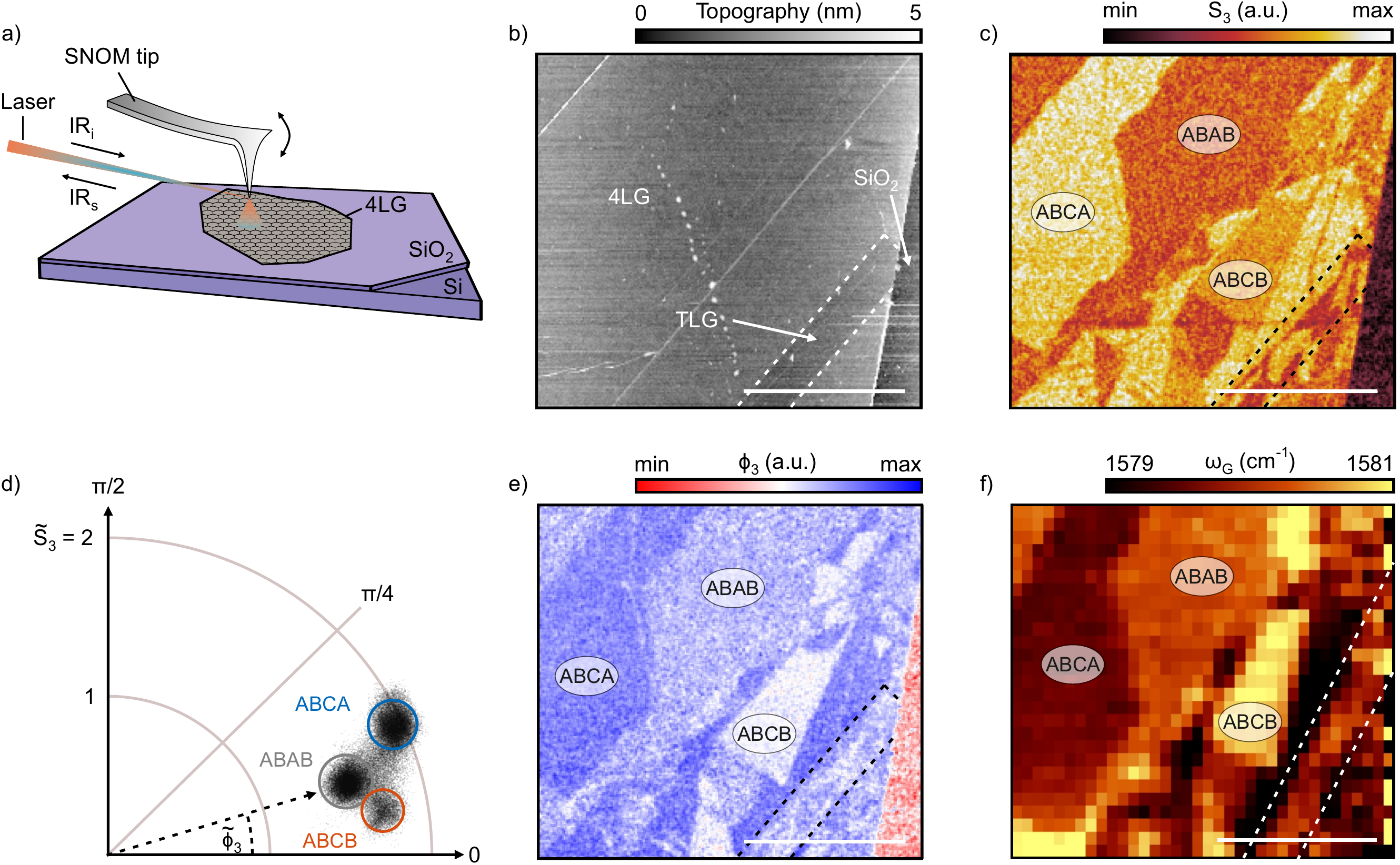}
        \caption{ \textbf{s-SNOM imaging of a 4LG flake.} a) Schematic of the s-SNOM set up with an infrared laser focused on the apex of the atomic force microscope tip. b) Topography image of a 4LG flake on SiO\textsubscript{2}. c) s-SNOM optical amplitude (S\textsubscript{3}) and e) s-SNOM's phase ($\Phi_3$) images revealing domains of different s-SNOM signal, obtained at 0.34~{eV}. Three domains on the 4LG with different amplitude values are indicated by ABCB, ABCA and ABAB. The dashed rectangle marks the position  of the TLG stripe (compare with b)). d) Scatter plot of normalized amplitude ($\tilde{S}$ = S\textsubscript{3}/S\textsubscript{3}(SiO\textsubscript{2})) and phase ($\tilde{\Phi}_3=\Phi_3-\Phi_{3}$(SiO\textsubscript{2})) of individual pixels obtained at 0.34 eV on the 4LG  plotted in a polar plot and referenced to the SiO\textsubscript{2}. The circles are a guide to the eye to identify the respective domains. f) Raman map of the G peak position. The three different domains marked in c) can also be identified. The dashed lines mark the border between 4LG and TLG (see panel b)). The scale bars corresponds to 5 µm.}
        \label{fig:1}
    \end{center}
\end{figure*}

Figure~\ref{fig:1}~a) shows the schematic of a scattering-type scanning near field optical microscope used to investigate the stacking order of 4LG flakes. Infrared light from a broadly tunable laser source is focused onto an atomic force microscope tip operated in tapping mode. 
From the back-scattered light, near-field signals at higher harmonics of the probe's tapping frequency are obtained. 
Here, we show third order optical amplitude (S\textsubscript{3}) and phase ($\Phi_3$) signals, which are recorded simultaneously to the topography (see Methods)~\cite{richards_near-field_2004}.  

In Figure~\ref{fig:1} b), the topography of a scanned 4LG flake (Flake 1) is shown. 
It reveals two segments of 4LG separated by a small stripe of trilayer graphene (TLG).
Except for two diagonal folds, the 4LG is mostly homogeneous.
The simultaneously recorded optical amplitude (S\textsubscript{3}) and phase images ($\Phi_3$) in Figures~\ref{fig:1} c) and e) are obtained at photon energies of 0.34~{eV}. 
In these images, the FLG flake can be clearly distinguished from the SiO\textsubscript{2} substrate.

In the large 4LG segment, distinct amplitude and phase values are present in differently sized domains across the flake.
The amplitude $\tilde{S}$=S\textsubscript{3}/S\textsubscript{3}(SiO\textsubscript{2}) and phase $\tilde{\Phi}_3=\Phi_3$-$\Phi_{3}$(SiO\textsubscript{2}) contrasts extracted from the 4LG area, referenced to the adjacent SiO\textsubscript{2}, are shown in a polar plot in Figure~\ref{fig:1} d).
Whereas pixel-to-pixel fluctuations are relatively large, three distinct clusters of different amplitude and phase response are identified. 
The different near-field responses originate from different optical conductivities across the 4LG, which the s-SNOM is sensitive to~\cite{kim_stacking_2015, wirth_tunable_2021}.
As the conductivities are connected to the electronic structure, we associate the distinct near-field responses with the three crystallographic stackings of 4LG~\cite{mak_electronic_2010} (ABAB, ABCA and ABCB).

The stacking assignment is supported by the different spectroscopic response at various infrared photon energies, which is also used to assign the domains in Figures \ref{fig:1} c)-f), and is discussed further in the text. 
Further evidence comes from confocal Raman spectroscopy of the few-layer graphene G, M, and 2D peaks.
A Raman map of the G peak position is shown in Figure~\ref{fig:1} f). 
The largest domains of each stacking sequence, as identified in the s-SNOM amplitude and phase images by the labels, similarly show three distinct peak positions. 
However, some smaller domains observed in Figure~\ref{fig:1} c) and e) cannot be fully resolved in the Raman map due to the diffraction limit.

We now turn to the characterization of the ABCB domains and discuss its fingerprints in Raman spectroscopy as well as in its infrared optical response. 

\subsection{Raman spectra}
\begin{figure}[!bht]
    \begin{center}
        \includegraphics[width=0.65\textwidth]{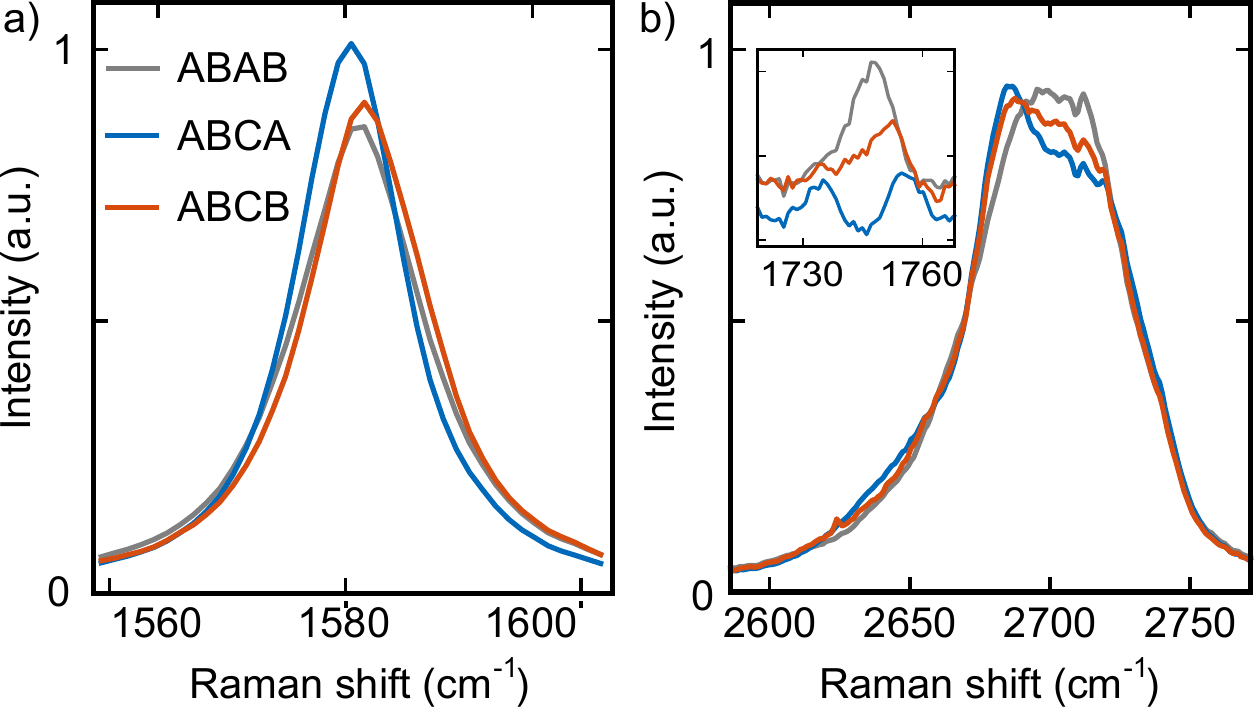}
        \caption{\textbf{Raman spectroscopy of the 4LG stackings.} Raman spectra of a) the G peaks and b) the 2D peaks of the three stacking orders  (normalized by their total intensity). The inset in b) shows the M peaks.}
        \label{fig:Raman}
    \end{center}
\end{figure}

Raman spectra of the G and 2D peak of the three stacking orders are shown in Figures~\ref{fig:Raman} a) and b), respectively (M peaks in the inset). 
Within individual samples, the G peak position and width show small changes between different domains, even though their absolute values depend on strain and doping~\cite{toporski_confocal_2018,Neumann2015Sep} and vary between samples. 
We observe the G-peak position of ABCA stacking to be red-shifted compared to ABAB, which is consistent with the shift observed in ABC and ABA stacked trilayer graphene~\cite{lui_imaging_2011}. 
For ABCB domains, we find the G peak to be at slightly higher energies compared to both other stackings. 
Notably, the full width at half maximum (FWHM) linewidth of the G peak is the smallest in the ABCA domains and largest in the ABAB domains.

The 2D and M peaks both originate from two-phonon processes and are thus sensitive to the electronic structure as well~\cite{Malard_2009, Cong_2011, torche_first-principles_2017}.
As a result, both peaks show distinct line shapes for all three stacking orders.  
Compared to the ABAB 2D peak, which is quite symmetric and featureless, the ABCA 2D peak shows a stronger asymmetry, a sharp feature  around 2680~{cm$^{-1}$} and a shoulder around 2640~{cm$^{-1}$}. 
These signatures are consistent with previously reported Raman spectra~\cite{lui_imaging_2011, nguyen_excitation_2014, geisenhof_anisotropic_2019, torche_first-principles_2017}.
The ABCB 2D peak appears like an interpolation of the two other peaks: it is more asymmetric than the ABAB peak, but shows a less pronounced feature on the low-energy side compared to the ABCA peak and no low-energy shoulder. 
The M-peaks \cite{Cong_2011} show a unique peak shape and position for each stacking, possibly making them the most reliable feature for domain identification despite their low intensity. 
The Raman spectra confirm the domain assignment by our s-SNOM measurements.
A detailed study of the Raman G-, 2D and and M-peaks of ABCB is beyond the scope of this paper.

\subsection{Optical conductivity}
\begin{figure*}[!bth]
    \begin{center}
        \includegraphics[width=1.0\textwidth]{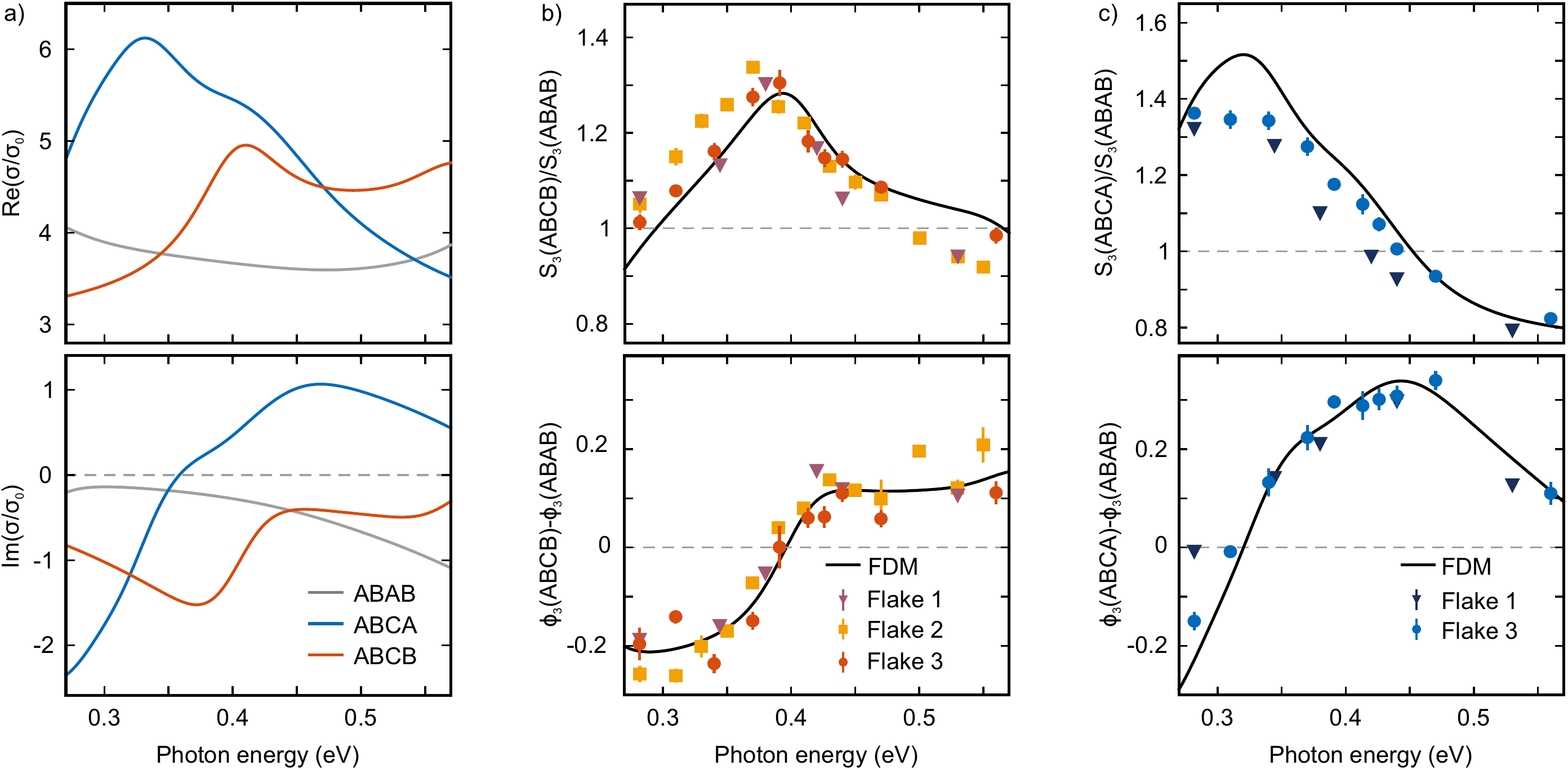}
        \caption{ \textbf{s-SNOM nano-spectroscopy of 4LG stackings}. a) Real- and imaginary part of the optical conductivity of the three different stacking orders obtained from tight binding calculations of 4LG for $\mu = 50$~{meV} with a phenomenologically chosen broadening of $\eta = 40$~{meV}. b) Third demodulation order amplitude S\textsubscript{3} (top) and phase $\Phi_3$ (bottom) data compared to theoretical s-SNOM spectra of ABCB referenced to ABAB for three different flakes. c) Third demodulation order amplitude (top) and phase (bottom) data compared to theoretical spectra for ABCA referenced to ABAB. Images of Flake 1 and Flake 2 can be found in the SI. The theoretical spectra are obtained from finite dipole model calculations.}
        \label{fig:2}
    \end{center}
\end{figure*}

To characterize the different optical conductivities following from the electronic structures, sequential nano-spectroscopy between 0.28~{eV} and 0.56~{eV} in conjunction with theoretical modelling  was conducted.
Figure~\ref{fig:2} a) shows the calculated optical conductivities $\sigma$ of the three crystallographic stackings ABAB, ABCA and ABCB.
The optical conductivities are obtained from tight binding calculations including nearest-neighbor interlayer- and nearest-neighbor intralayer hopping with hopping energies of $\gamma_{\mathrm{inter}} = 0.39$\,eV and $\gamma_{\mathrm{intra}} = 3.16$\,eV, respectively (see SI S4).
This Hamiltonian is optimized to reproduce the band structure at the energies of the s-SNOM measurements \cite{aoki_dependence_2007, mak_electronic_2010}.
The chemical potential ($\mu = 50$~{meV}) and a phenomenological broadening ($\eta = 40$~{meV}) are chosen to achieve a good match with the s-SNOM data of Flake 3 (see SI S3).

Real- and imaginary part of the conductivity of ABAB are featureless and almost flat between 0.28 and 0.56~{eV}. Accordingly, we expect a homogeneous amplitude ($S_3$) and phase ($\Phi_3$) response for ABAB when compared to ABCA and ABCB, because these reproduce the characteristic features of real and imaginary part of the conductivity~\cite{wirth_tunable_2021}.

In total, three different 4LG flakes showing three different domains at 0.34~{eV} were investigated (Flake 1 in Figure~\ref{fig:1}; Flake 2 and Flake 3 in SI S1) by s-SNOM. 
The amplitude and phase contrasts are evaluated line-wise, because of a phase drift, and referenced to the adjacent ABAB values. 
This referencing overcomes the strong tip dependence of the s-SNOM response of these thin films and makes the results from the three flakes quantitatively comparable.

Amplitude and phase spectra of ABCB- and ABCA-domains, referenced to the ABAB-domains, are shown in Figures~\ref{fig:2} b) and c). 
Note, that Flake 1, 2 and 3 refer to three independently investigated 4LG flakes (see SI S1).
The theoretical spectra were obtained from the optical conductivities using a multilayer extension of the finite-dipole model (FDM) \cite{hauer_quasi-analytical_2012}, in which the graphene layers are taken into account as infinitesimal thin interfaces with a conductivity obtained from tight binding calculations (for Details see SI S2). 
Both amplitude and phase data reproduce the energy-dependence of the calculated s-SNOM contrast well, constituting the main assignment of the domains in Figure~\ref{fig:1} c).

The normalized amplitude data sets S\textsubscript{3}(ABCB)/S\textsubscript{3}(ABAB) exhibit a peak at 0.38~{eV} (Figure~\ref{fig:2} b)).
The corresponding phase contrast $\Phi_3$(ABCB)-$\Phi_{3}$(ABAB) increases with photon energy and has a zero crossing around $0.4$\,eV. 
For comparison, amplitude and phase contrast data of domain ABCA are plotted in Figure~\ref{fig:2} c). 
Similar domains were investigated on two of the three flakes. 
The ABCA spectral amplitude and phase data are clearly distinct from the response of ABCB: at low energies down to 0.28~{eV} the amplitude contrast is much larger than in b) while above, it is monotonically decreasing with  photon energy to values below one above $0.45$ eV.
The experimental data for both stackings, referenced to ABAB, quantitatively match the calculation at almost all measured photon energies.
This shows that the optical response in the entire IR region from 0.28~{eV} to 0.56~{eV} can be used to distinguish the three stacking orders.
Thus, the detection of ABCB domains in s-SNOM is not restricted to our set-up employing a tunable laser source, but could also be achieved with a more commonly available Helium-Neon laser, which can be operated at 0.366 eV, and has been used to distinguish between TLG and FLG stackings in s-SNOM ~\cite{kim_stacking_2015}.

\subsection{Abundance and stability of ABCB domains}
Theoretical calculations suggest that ABCB domains are metastable at a higher total energy than ABAB and ABCA~\cite{aoki_dependence_2007}, whereas in published experimental work, the stacking had so far not been observed.
In total, we have scanned an area of 40000 \textmu m$^2$ of exfoliated tetralayer graphene and found ABCB domains to make up approximately 2\% of the total area (ABAB to ABCA were around 71\% and 27\%, respectively).
The largest domain was approximately $100$ \textmu m$^2$ in size, whereas most domains are significantly smaller.
This might explain why, so far, these domains have been overlooked in published work.
Based on the s-SNOM and Raman results described in our current work, in the meantime we were also able to unambiguously verify the existence of ABCB by performing Fourier transform infrared micro-spectroscopy (FTIR) on flakes with largest domains, yielding distinct IR reflectance spectra for the three possible stackings of 4LG (See SI Section S5 and Figure S4).
So far, it is not known if the size and abundance of ABCB domains after exfoliation depends on the bulk crystal, but we observe that it varies between exfoliations. 
This might be due to the applied lateral strain \cite{Yang2019, Nery2020} and needs to be investigated more closely.

The ABCB domains were stable over the course of several weeks at ambient conditions as well as when subjected to s-SNOM and Raman measurements at moderate laser powers. 
At higher laser power, we observed the shrinkage of some domains (for details see SI S1) which has also been observed for metastable rhombohedral TLG \cite{zhang_light-induced_2020}.

\section{Discussion}
Our experimental results reveal the existence of ABCB stacked tetralayer graphene domains.
This observation is enabled by addressing the low-energy electronic structure through the optical conductivity with a tunable laser between 0.28 and 0.56~{eV} in an s-SNOM setup. The optical conductivity has unique fingerprints arising from characteristic interband transitions for each stacking which manifest themselves in peaks in the normalized scattering amplitude and phase spectra.
These features show a broadening of approximately $40$~{meV}, which is below reported values of 50~meV in comparable FTIR studies~\cite{mak_evolution_2010}.
By referencing the adjacent domains to each other, we achieved excellent agreement between theory and experiment, establishing s-SNOM as a semi-quantitative tool for nanoscale spectroscopy over a wide energy range in the near-infrared spectral region. 

From the stability and abundance of our measured samples, we conclude that ABCB domains are metastable at room temperature, although, the least abundant occurring domains after tape exfoliation. This agrees with theoretical predictions where the energy barrier between ABCB and ABAB is expected to be lower when compared to ABCA and ABCB~\cite{aoki_dependence_2007}. 

The new stacking is of high interest, because it is the thinnest naturally occurring FLG stacking with an intrinsic bandgap~\cite{latil_charge_2006, aoki_dependence_2007, min_electronic_2008} and its low energy physics is dominated by valley flat bands which might promote electron-electron interactions favoring correlated states, such as magnetism and unconventional superconductivity. The intricate nature of these states is an interesting avenue for further theoretical~\cite{InPrep} and experimental studies. 
Furthermore, playing to the unique intrinsic narrow band-gap insulating behavior of ABCB graphene, one faces the natural question whether correlated or excitonic physics might play a prominent role.
Studying these questions is an intriguing avenue of future research, for which our characterisation of the optical response of ABCB graphene will serve as a convenient starting point.

\section{Methods}
\paragraph*{Exfoliation}
We obtain tetralayer graphene flakes by exfoliation of commercially available graphite crystals (Graphenium Flakes, NGS Naturgraphit GmbH).
Mechanical cleaving is performed using a dicing tape (1008R, Ultron Systems), which is then pressed onto silicon wafers with an oxide thickness of $90$~{nm}.
All processes are performed at room temperature under ambient conditions. 
No additionally cleaning steps were performed.
Suitable flakes were identified via their optical contrast using a standard optical microscope. 
\paragraph*{s-SNOM}
We use a commercially available s-SNOM (NeaSNOM, neaspec GmbH) at photon energies between 0.28 to 0.56~{eV} for sequential nano-spectroscopy on 4LG in combination with a LN2 cooled InSb detector (Infrared Associates). The s-SNOM is operated in pseudo-heterodyne detection mode, to record amplitude and phase simultaneously. The laser source is a commercially available tunable OPO/OPA laser system (Alpha Module, Stuttgart Instruments) with an energy resolution of $6$~{meV}, which can address the energy range from 0.27 to 0.9~{eV}. The laser source emits pulses with a pulse length of 1~{ps} and has a repetition rate of 42 MHz. The s-SNOM is operated at tapping frequencies of 220-270~{kHz} at tapping amplitudes of 50-60~{nm}. We extracted the signal from the third demodulation order amplitude and phase. For single imaging with s-SNOM in pseudo-heterodyne detection mode, this pulse width is sufficient in the investigated energy range as recently shown \cite{wirth_tunable_2021, lu_observing_nodate}.

\paragraph*{Raman spectroscopy}
Raman maps were taken using a 532~{nm} laser focused down to a spotsize of approximately 500~{nm}. The Raman spectra shown in Figure~\ref{fig:2}, were obtained by averaging the Raman maps within the individual stacking domains. The excitation power was mostly kept to 1.5~{mW} or below, as higher laser powers occasionally resulted in shrinkage of ABCA and ABCB domains.

\begin{acknowledgement}
This work was supported by the Excellence Initiative of the
German federal and state governments, the Ministry of Innovation of North Rhine-Westphalia and the Deutsche Forschungsgemeinschaft.
KGW, DS and TT acknowledge support from the Deutsche Forschungsgemeinschaft (DFG) within the collaborative research center SFB~917 and within Grant Agreement No. TA 848/7-1.
AR, BB, CS, and LW acknowledge support from the European Union’s Horizon 2020 research and innovation programme under grant agreement No. 881603 (Graphene Flagship), the Deutsche Forschungsgemeinschaft (DFG, German Research Foundation) under Germany’s Excellence Strategy - Cluster of Excellence Matter and Light for Quantum Computing (ML4Q) EXC 2004/1 - 390534769, through DFG (BE 2441/9-1), and the FLAG-ERA grant TATTOOS, by the Deutsche Forschungsgemeinschaft (DFG, German Research Foundation) - 437214324.
JBH, LK, AF and DMK acknowledge funding by the Deutsche Forschungsgemeinschaft (DFG, German Research Foundation) under RTG 1995, within the Priority Program SPP 2244 ``2DMP'' and under Germany’s Excellence Strategy - Cluster of Excellence Matter and Light for Quantum Computing (ML4Q) EXC 2004/1 - 390534769. DMK acknowledges support by the Max Planck-New York City Center for Nonequilibrium Quantum Phenomena.
We acknowledge computational resources provided by the Max Planck Computing and Data Facility and RWTH Aachen University under project number rwth0742 and rwth0716.
\end{acknowledgement}

\section{Author Contributions}
KGW, JBH, LW, DMK, BB, CS and TT conceived the project.
AR and HK fabricated the samples.
KGW and DS performed the s-SNOM experiments and theoretical contrast calculations.
LC conducted the FTIR measurements.
AR, HK and LW carried out the Raman measurements. 
LW and KGW analysed the experimental data. 
JBH, LK and AF carried out the theoretical calculation. 
All authors contributed to writing the manuscript.

\section{Competing interests}
The authors declare no competing interests.

\nocite{cvitkovic_analytical_2007, zhan_transfer_2013, kubo_statistical-mechanical_1957, Rubio_flg_Tb}

\bibliography{literature4LG}

\providecommand{\latin}[1]{#1}
\makeatletter
\providecommand{\doi}
  {\begingroup\let\do\@makeother\dospecials
  \catcode`\{=1 \catcode`\}=2 \doi@aux}
\providecommand{\doi@aux}[1]{\endgroup\texttt{#1}}
\makeatother
\providecommand*\mcitethebibliography{\thebibliography}
\csname @ifundefined\endcsname{endmcitethebibliography}
  {\let\endmcitethebibliography\endthebibliography}{}
\begin{mcitethebibliography}{51}
\providecommand*\natexlab[1]{#1}
\providecommand*\mciteSetBstSublistMode[1]{}
\providecommand*\mciteSetBstMaxWidthForm[2]{}
\providecommand*\mciteBstWouldAddEndPuncttrue
  {\def\EndOfBibitem{\unskip.}}
\providecommand*\mciteBstWouldAddEndPunctfalse
  {\let\EndOfBibitem\relax}
\providecommand*\mciteSetBstMidEndSepPunct[3]{}
\providecommand*\mciteSetBstSublistLabelBeginEnd[3]{}
\providecommand*\EndOfBibitem{}
\mciteSetBstSublistMode{f}
\mciteSetBstMaxWidthForm{subitem}{(\alph{mcitesubitemcount})}
\mciteSetBstSublistLabelBeginEnd
  {\mcitemaxwidthsubitemform\space}
  {\relax}
  {\relax}

\bibitem[Zhang \latin{et~al.}(2005)Zhang, Tan, Stormer, and Kim]{Zhang2005Nov}
Zhang,~Y.; Tan,~Y.-W.; Stormer,~H.~L.; Kim,~P. {Experimental observation of the
  quantum Hall effect and Berry's phase in graphene}. \emph{Nature}
  \textbf{2005}, \emph{438}, 201--204\relax
\mciteBstWouldAddEndPuncttrue
\mciteSetBstMidEndSepPunct{\mcitedefaultmidpunct}
{\mcitedefaultendpunct}{\mcitedefaultseppunct}\relax
\EndOfBibitem
\bibitem[Novoselov \latin{et~al.}(2006)Novoselov, McCann, Morozov, Fal{'}ko,
  Katsnelson, Zeitler, Jiang, Schedin, and Geim]{Novoselov2006Mar}
Novoselov,~K.~S.; McCann,~E.; Morozov,~S.~V.; Fal{'}ko,~V.~I.;
  Katsnelson,~M.~I.; Zeitler,~U.; Jiang,~D.; Schedin,~F.; Geim,~A.~K.
  {Unconventional quantum Hall effect and Berry{'}s phase of 2{$\pi$} in
  bilayer graphene}. \emph{Nat. Phys.} \textbf{2006}, \emph{2}, 177--180\relax
\mciteBstWouldAddEndPuncttrue
\mciteSetBstMidEndSepPunct{\mcitedefaultmidpunct}
{\mcitedefaultendpunct}{\mcitedefaultseppunct}\relax
\EndOfBibitem
\bibitem[Zhou \latin{et~al.}(2021)Zhou, Saito, Cohen, Huynh, Patterson, Yang,
  Taniguchi, Watanabe, and Young]{zhou_isospin_2021}
Zhou,~H.; Saito,~Y.; Cohen,~L.; Huynh,~W.; Patterson,~C.~L.; Yang,~F.;
  Taniguchi,~T.; Watanabe,~K.; Young,~A.~F. Isospin magnetism and spin-triplet
  superconductivity in {Bernal} bilayer graphene. \emph{arXiv:2110.11317
  [cond-mat]} \textbf{2021}, \relax
\mciteBstWouldAddEndPunctfalse
\mciteSetBstMidEndSepPunct{\mcitedefaultmidpunct}
{}{\mcitedefaultseppunct}\relax
\EndOfBibitem
\bibitem[Zhou \latin{et~al.}(2021)Zhou, Xie, Taniguchi, Watanabe, and
  Young]{zhou_superconductivity_2021}
Zhou,~H.; Xie,~T.; Taniguchi,~T.; Watanabe,~K.; Young,~A.~F. Superconductivity
  in rhombohedral trilayer graphene. \emph{Nature} \textbf{2021}, \emph{598},
  434--438\relax
\mciteBstWouldAddEndPuncttrue
\mciteSetBstMidEndSepPunct{\mcitedefaultmidpunct}
{\mcitedefaultendpunct}{\mcitedefaultseppunct}\relax
\EndOfBibitem
\bibitem[Zhou \latin{et~al.}(2021)Zhou, Xie, Ghazaryan, Holder, Ehrets,
  Spanton, Taniguchi, Watanabe, Berg, Serbyn, and Young]{zhou_half-_2021}
Zhou,~H.; Xie,~T.; Ghazaryan,~A.; Holder,~T.; Ehrets,~J.~R.; Spanton,~E.~M.;
  Taniguchi,~T.; Watanabe,~K.; Berg,~E.; Serbyn,~M.; Young,~A.~F. Half- and
  quarter-metals in rhombohedral trilayer graphene. \emph{Nature}
  \textbf{2021}, \emph{598}, 429--433\relax
\mciteBstWouldAddEndPuncttrue
\mciteSetBstMidEndSepPunct{\mcitedefaultmidpunct}
{\mcitedefaultendpunct}{\mcitedefaultseppunct}\relax
\EndOfBibitem
\bibitem[Kerelsky \latin{et~al.}(2021)Kerelsky, Rubio-Verdú, Xian, Kennes,
  Halbertal, Finney, Song, Turkel, Wang, Watanabe, Taniguchi, Hone, Dean,
  Basov, Rubio, and Pasupathy]{kerelsky_moireless_2021}
Kerelsky,~A.; Rubio-Verdú,~C.; Xian,~L.; Kennes,~D.~M.; Halbertal,~D.;
  Finney,~N.; Song,~L.; Turkel,~S.; Wang,~L.; Watanabe,~K.; Taniguchi,~T.;
  Hone,~J.; Dean,~C.; Basov,~D.~N.; Rubio,~A.; Pasupathy,~A.~N. Moiréless
  correlations in {ABCA} graphene. \emph{Proceedings of the National Academy of
  Sciences} \textbf{2021}, \emph{118}, e2017366118\relax
\mciteBstWouldAddEndPuncttrue
\mciteSetBstMidEndSepPunct{\mcitedefaultmidpunct}
{\mcitedefaultendpunct}{\mcitedefaultseppunct}\relax
\EndOfBibitem
\bibitem[Balents \latin{et~al.}(2020)Balents, Dean, Efetov, and
  Young]{Balents2020}
Balents,~L.; Dean,~C.~R.; Efetov,~D.~K.; Young,~A.~F. Superconductivity and
  strong correlations in moir{\'e} flat bands. \emph{Nature Physics}
  \textbf{2020}, \emph{16}, 725--733\relax
\mciteBstWouldAddEndPuncttrue
\mciteSetBstMidEndSepPunct{\mcitedefaultmidpunct}
{\mcitedefaultendpunct}{\mcitedefaultseppunct}\relax
\EndOfBibitem
\bibitem[Kennes \latin{et~al.}(2021)Kennes, Claassen, Xian, Georges, Millis,
  Hone, Dean, Basov, Pasupathy, and Rubio]{kennes_moire_2021}
Kennes,~D.~M.; Claassen,~M.; Xian,~L.; Georges,~A.; Millis,~A.~J.; Hone,~J.;
  Dean,~C.~R.; Basov,~D.~N.; Pasupathy,~A.~N.; Rubio,~A. Moiré
  heterostructures as a condensed-matter quantum simulator. \emph{Nature
  Physics} \textbf{2021}, \emph{17}, 155--163\relax
\mciteBstWouldAddEndPuncttrue
\mciteSetBstMidEndSepPunct{\mcitedefaultmidpunct}
{\mcitedefaultendpunct}{\mcitedefaultseppunct}\relax
\EndOfBibitem
\bibitem[Cao \latin{et~al.}(2018)Cao, Fatemi, Fang, Watanabe, Taniguchi,
  Kaxiras, and Jarillo-Herrero]{cao_unconventional_2018}
Cao,~Y.; Fatemi,~V.; Fang,~S.; Watanabe,~K.; Taniguchi,~T.; Kaxiras,~E.;
  Jarillo-Herrero,~P. Unconventional superconductivity in magic-angle graphene
  superlattices. \emph{Nature} \textbf{2018}, \emph{556}, 43--50\relax
\mciteBstWouldAddEndPuncttrue
\mciteSetBstMidEndSepPunct{\mcitedefaultmidpunct}
{\mcitedefaultendpunct}{\mcitedefaultseppunct}\relax
\EndOfBibitem
\bibitem[Liu \latin{et~al.}(2020)Liu, Hao, Khalaf, Lee, Ronen, Yoo,
  Haei~Najafabadi, Watanabe, Taniguchi, Vishwanath, and Kim]{liu_tunable_2020}
Liu,~X.; Hao,~Z.; Khalaf,~E.; Lee,~J.~Y.; Ronen,~Y.; Yoo,~H.;
  Haei~Najafabadi,~D.; Watanabe,~K.; Taniguchi,~T.; Vishwanath,~A.; Kim,~P.
  Tunable spin-polarized correlated states in twisted double bilayer graphene.
  \emph{Nature} \textbf{2020}, \emph{583}, 221--225\relax
\mciteBstWouldAddEndPuncttrue
\mciteSetBstMidEndSepPunct{\mcitedefaultmidpunct}
{\mcitedefaultendpunct}{\mcitedefaultseppunct}\relax
\EndOfBibitem
\bibitem[Shen \latin{et~al.}(2020)Shen, Chu, Wu, Li, Wang, Zhao, Tang, Liu,
  Tian, Watanabe, Taniguchi, Yang, Meng, Shi, Yazyev, and
  Zhang]{shen_correlated_2020}
Shen,~C.; Chu,~Y.; Wu,~Q.; Li,~N.; Wang,~S.; Zhao,~Y.; Tang,~J.; Liu,~J.;
  Tian,~J.; Watanabe,~K.; Taniguchi,~T.; Yang,~R.; Meng,~Z.~Y.; Shi,~D.;
  Yazyev,~O.~V.; Zhang,~G. Correlated states in twisted double bilayer
  graphene. \emph{Nature Physics} \textbf{2020}, \emph{16}, 520--525\relax
\mciteBstWouldAddEndPuncttrue
\mciteSetBstMidEndSepPunct{\mcitedefaultmidpunct}
{\mcitedefaultendpunct}{\mcitedefaultseppunct}\relax
\EndOfBibitem
\bibitem[Aoki and Amawashi(2007)Aoki, and Amawashi]{aoki_dependence_2007}
Aoki,~M.; Amawashi,~H. Dependence of band structures on stacking and field in
  layered graphene. \emph{Solid State Communications} \textbf{2007},
  \emph{142}, 123--127\relax
\mciteBstWouldAddEndPuncttrue
\mciteSetBstMidEndSepPunct{\mcitedefaultmidpunct}
{\mcitedefaultendpunct}{\mcitedefaultseppunct}\relax
\EndOfBibitem
\bibitem[Zhang \latin{et~al.}(2020)Zhang, Han, Peng, Yang, Yuan, Li, Chen, Xu,
  Liu, Zhu, Cao, Han, Dai, Zhu, Qin, and Novoselov]{zhang_light-induced_2020}
Zhang,~J.; Han,~J.; Peng,~G.; Yang,~X.; Yuan,~X.; Li,~Y.; Chen,~J.; Xu,~W.;
  Liu,~K.; Zhu,~Z.; Cao,~W.; Han,~Z.; Dai,~J.; Zhu,~M.; Qin,~S.;
  Novoselov,~K.~S. Light-induced irreversible structural phase transition in
  trilayer graphene. \emph{Light: Science \& Applications} \textbf{2020},
  \emph{9}, 174\relax
\mciteBstWouldAddEndPuncttrue
\mciteSetBstMidEndSepPunct{\mcitedefaultmidpunct}
{\mcitedefaultendpunct}{\mcitedefaultseppunct}\relax
\EndOfBibitem
\bibitem[Lui \latin{et~al.}(2011)Lui, Li, Chen, Klimov, Brus, and
  Heinz]{lui_imaging_2011}
Lui,~C.~H.; Li,~Z.; Chen,~Z.; Klimov,~P.~V.; Brus,~L.~E.; Heinz,~T.~F. Imaging
  {Stacking} {Order} in {Few}-{Layer} {Graphene}. \emph{Nano Letters}
  \textbf{2011}, \emph{11}, 164--169\relax
\mciteBstWouldAddEndPuncttrue
\mciteSetBstMidEndSepPunct{\mcitedefaultmidpunct}
{\mcitedefaultendpunct}{\mcitedefaultseppunct}\relax
\EndOfBibitem
\bibitem[Latil and Henrard(2006)Latil, and Henrard]{latil_charge_2006}
Latil,~S.; Henrard,~L. Charge {Carriers} in {Few}-{Layer} {Graphene} {Films}.
  \emph{Physical Review Letters} \textbf{2006}, \emph{97}, 036803\relax
\mciteBstWouldAddEndPuncttrue
\mciteSetBstMidEndSepPunct{\mcitedefaultmidpunct}
{\mcitedefaultendpunct}{\mcitedefaultseppunct}\relax
\EndOfBibitem
\bibitem[Min and MacDonald(2008)Min, and MacDonald]{min_electronic_2008}
Min,~H.; MacDonald,~A.~H. Electronic {Structure} of {Multilayer} {Graphene}.
  \emph{Progress of Theoretical Physics Supplement} \textbf{2008}, \emph{176},
  227--252\relax
\mciteBstWouldAddEndPuncttrue
\mciteSetBstMidEndSepPunct{\mcitedefaultmidpunct}
{\mcitedefaultendpunct}{\mcitedefaultseppunct}\relax
\EndOfBibitem
\bibitem[Mak \latin{et~al.}(2010)Mak, Sfeir, Misewich, and
  Heinz]{mak_evolution_2010}
Mak,~K.~F.; Sfeir,~M.~Y.; Misewich,~J.~A.; Heinz,~T.~F. The evolution of
  electronic structure in few-layer graphene revealed by optical spectroscopy.
  \emph{Proceedings of the National Academy of Sciences} \textbf{2010},
  \emph{107}, 14999--15004\relax
\mciteBstWouldAddEndPuncttrue
\mciteSetBstMidEndSepPunct{\mcitedefaultmidpunct}
{\mcitedefaultendpunct}{\mcitedefaultseppunct}\relax
\EndOfBibitem
\bibitem[Mak \latin{et~al.}(2010)Mak, Shan, and Heinz]{mak_electronic_2010}
Mak,~K.~F.; Shan,~J.; Heinz,~T.~F. Electronic {Structure} of {Few}-{Layer}
  {Graphene}: {Experimental} {Demonstration} of {Strong} {Dependence} on
  {Stacking} {Sequence}. \emph{Physical Review Letters} \textbf{2010},
  \emph{104}, 176404\relax
\mciteBstWouldAddEndPuncttrue
\mciteSetBstMidEndSepPunct{\mcitedefaultmidpunct}
{\mcitedefaultendpunct}{\mcitedefaultseppunct}\relax
\EndOfBibitem
\bibitem[Lui \latin{et~al.}(2011)Lui, Li, Mak, Cappelluti, and
  Heinz]{lui_observation_2011}
Lui,~C.~H.; Li,~Z.; Mak,~K.~F.; Cappelluti,~E.; Heinz,~T.~F. Observation of an
  electrically tunable band gap in trilayer graphene. \emph{Nature Physics}
  \textbf{2011}, \emph{7}, 944--947\relax
\mciteBstWouldAddEndPuncttrue
\mciteSetBstMidEndSepPunct{\mcitedefaultmidpunct}
{\mcitedefaultendpunct}{\mcitedefaultseppunct}\relax
\EndOfBibitem
\bibitem[Berciaud \latin{et~al.}(2014)Berciaud, Potemski, and
  Faugeras]{berciaud_probing_2014}
Berciaud,~S.; Potemski,~M.; Faugeras,~C. Probing {Electronic} {Excitations} in
  {Mono}- to {Pentalayer} {Graphene} by {Micro} {Magneto}-{Raman}
  {Spectroscopy}. \emph{Nano Letters} \textbf{2014}, \emph{14}, 4548--4553,
  Publisher: American Chemical Society\relax
\mciteBstWouldAddEndPuncttrue
\mciteSetBstMidEndSepPunct{\mcitedefaultmidpunct}
{\mcitedefaultendpunct}{\mcitedefaultseppunct}\relax
\EndOfBibitem
\bibitem[Henni \latin{et~al.}(2016)Henni, Ojeda~Collado, Nogajewski, Molas,
  Usaj, Balseiro, Orlita, Potemski, and Faugeras]{henni_rhombohedral_2016}
Henni,~Y.; Ojeda~Collado,~H.~P.; Nogajewski,~K.; Molas,~M.~R.; Usaj,~G.;
  Balseiro,~C.~A.; Orlita,~M.; Potemski,~M.; Faugeras,~C. Rhombohedral
  {Multilayer} {Graphene}: {A} {Magneto}-{Raman} {Scattering} {Study}.
  \emph{Nano Letters} \textbf{2016}, \emph{16}, 3710--3716\relax
\mciteBstWouldAddEndPuncttrue
\mciteSetBstMidEndSepPunct{\mcitedefaultmidpunct}
{\mcitedefaultendpunct}{\mcitedefaultseppunct}\relax
\EndOfBibitem
\bibitem[Taubner \latin{et~al.}(2003)Taubner, Hillenbrand, and
  Keilmann]{taubner_performance_2003}
Taubner,~T.; Hillenbrand,~R.; Keilmann,~F. Performance of visible and
  mid-infrared scattering-type near-field optical microscopes. \emph{Journal of
  Microscopy} \textbf{2003}, \emph{210}, 311--314\relax
\mciteBstWouldAddEndPuncttrue
\mciteSetBstMidEndSepPunct{\mcitedefaultmidpunct}
{\mcitedefaultendpunct}{\mcitedefaultseppunct}\relax
\EndOfBibitem
\bibitem[Fei \latin{et~al.}(2012)Fei, Rodin, Andreev, Bao, McLeod, Wagner,
  Zhang, Zhao, Thiemens, Dominguez, Fogler, Neto, Lau, Keilmann, and
  Basov]{fei_gate-tuning_2012}
Fei,~Z.; Rodin,~A.~S.; Andreev,~G.~O.; Bao,~W.; McLeod,~A.~S.; Wagner,~M.;
  Zhang,~L.~M.; Zhao,~Z.; Thiemens,~M.; Dominguez,~G.; Fogler,~M.~M.; Neto,~A.
  H.~C.; Lau,~C.~N.; Keilmann,~F.; Basov,~D.~N. Gate-tuning of graphene
  plasmons revealed by infrared nano-imaging. \emph{Nature} \textbf{2012},
  \emph{487}, 82--85\relax
\mciteBstWouldAddEndPuncttrue
\mciteSetBstMidEndSepPunct{\mcitedefaultmidpunct}
{\mcitedefaultendpunct}{\mcitedefaultseppunct}\relax
\EndOfBibitem
\bibitem[Chen \latin{et~al.}(2012)Chen, Badioli, Alonso-González,
  Thongrattanasiri, Huth, Osmond, Spasenović, Centeno, Pesquera, Godignon,
  Zurutuza~Elorza, Camara, de~Abajo, Hillenbrand, and
  Koppens]{chen_optical_2012}
Chen,~J.; Badioli,~M.; Alonso-González,~P.; Thongrattanasiri,~S.; Huth,~F.;
  Osmond,~J.; Spasenović,~M.; Centeno,~A.; Pesquera,~A.; Godignon,~P.;
  Zurutuza~Elorza,~A.; Camara,~N.; de~Abajo,~F. J.~G.; Hillenbrand,~R.;
  Koppens,~F. H.~L. Optical nano-imaging of gate-tunable graphene plasmons.
  \emph{Nature} \textbf{2012}, \emph{487}, 77--81\relax
\mciteBstWouldAddEndPuncttrue
\mciteSetBstMidEndSepPunct{\mcitedefaultmidpunct}
{\mcitedefaultendpunct}{\mcitedefaultseppunct}\relax
\EndOfBibitem
\bibitem[Fei \latin{et~al.}(2013)Fei, Rodin, Gannett, Dai, Regan, Wagner, Liu,
  McLeod, Dominguez, Thiemens, Castro~Neto, Keilmann, Zettl, Hillenbrand,
  Fogler, and Basov]{fei_electronic_2013}
Fei,~Z.; Rodin,~A.~S.; Gannett,~W.; Dai,~S.; Regan,~W.; Wagner,~M.; Liu,~M.~K.;
  McLeod,~A.~S.; Dominguez,~G.; Thiemens,~M.; Castro~Neto,~A.~H.; Keilmann,~F.;
  Zettl,~A.; Hillenbrand,~R.; Fogler,~M.~M.; Basov,~D.~N. Electronic and
  plasmonic phenomena at graphene grain boundaries. \emph{Nature
  Nanotechnology} \textbf{2013}, \emph{8}, 821--825\relax
\mciteBstWouldAddEndPuncttrue
\mciteSetBstMidEndSepPunct{\mcitedefaultmidpunct}
{\mcitedefaultendpunct}{\mcitedefaultseppunct}\relax
\EndOfBibitem
\bibitem[Jiang \latin{et~al.}(2016)Jiang, Shi, Zeng, Wang, Kang, Joshi, Jin,
  Ju, Kim, Lyu, Shen, Crommie, Gao, and Wang]{jiang_soliton-dependent_2016}
Jiang,~L.; Shi,~Z.; Zeng,~B.; Wang,~S.; Kang,~J.-H.; Joshi,~T.; Jin,~C.;
  Ju,~L.; Kim,~J.; Lyu,~T.; Shen,~Y.-R.; Crommie,~M.; Gao,~H.-J.; Wang,~F.
  Soliton-dependent plasmon reflection at bilayer graphene domain walls.
  \emph{Nature Materials} \textbf{2016}, \emph{15}, 840--844\relax
\mciteBstWouldAddEndPuncttrue
\mciteSetBstMidEndSepPunct{\mcitedefaultmidpunct}
{\mcitedefaultendpunct}{\mcitedefaultseppunct}\relax
\EndOfBibitem
\bibitem[Jiang \latin{et~al.}(2017)Jiang, Ni, Addison, Shi, Liu, Zhao, Kim,
  Mele, Basov, and Fogler]{jiang_plasmon_2017}
Jiang,~B.-Y.; Ni,~G.-X.; Addison,~Z.; Shi,~J.~K.; Liu,~X.; Zhao,~S. Y.~F.;
  Kim,~P.; Mele,~E.~J.; Basov,~D.~N.; Fogler,~M.~M. Plasmon {Reflections} by
  {Topological} {Electronic} {Boundaries} in {Bilayer} {Graphene}. \emph{Nano
  Letters} \textbf{2017}, \emph{17}, 7080--7085\relax
\mciteBstWouldAddEndPuncttrue
\mciteSetBstMidEndSepPunct{\mcitedefaultmidpunct}
{\mcitedefaultendpunct}{\mcitedefaultseppunct}\relax
\EndOfBibitem
\bibitem[Jiang \latin{et~al.}(2018)Jiang, Wang, Shi, Jin, Utama, Zhao, Shen,
  Gao, Zhang, and Wang]{jiang_manipulation_2018}
Jiang,~L.; Wang,~S.; Shi,~Z.; Jin,~C.; Utama,~M. I.~B.; Zhao,~S.; Shen,~Y.-R.;
  Gao,~H.-J.; Zhang,~G.; Wang,~F. Manipulation of domain-wall solitons in bi-
  and trilayer graphene. \emph{Nature Nanotechnology} \textbf{2018}, \emph{13},
  204--208\relax
\mciteBstWouldAddEndPuncttrue
\mciteSetBstMidEndSepPunct{\mcitedefaultmidpunct}
{\mcitedefaultendpunct}{\mcitedefaultseppunct}\relax
\EndOfBibitem
\bibitem[Sunku \latin{et~al.}(2018)Sunku, Ni, Jiang, Yoo, Sternbach, McLeod,
  Stauber, Xiong, Taniguchi, Watanabe, Kim, Fogler, and
  Basov]{sunku_photonic_2018}
Sunku,~S.~S.; Ni,~G.~X.; Jiang,~B.~Y.; Yoo,~H.; Sternbach,~A.; McLeod,~A.~S.;
  Stauber,~T.; Xiong,~L.; Taniguchi,~T.; Watanabe,~K.; Kim,~P.; Fogler,~M.~M.;
  Basov,~D.~N. Photonic crystals for nano-light in moiré graphene
  superlattices. \emph{Science} \textbf{2018}, \emph{362}, 1153--1156\relax
\mciteBstWouldAddEndPuncttrue
\mciteSetBstMidEndSepPunct{\mcitedefaultmidpunct}
{\mcitedefaultendpunct}{\mcitedefaultseppunct}\relax
\EndOfBibitem
\bibitem[Halbertal \latin{et~al.}(2021)Halbertal, Finney, Sunku, Kerelsky,
  Rubio-Verdú, Shabani, Xian, Carr, Chen, Zhang, and
  et~al.]{halbertal-2021-metrology}
Halbertal,~D.; Finney,~N.~R.; Sunku,~S.~S.; Kerelsky,~A.; Rubio-Verdú,~C.;
  Shabani,~S.; Xian,~L.; Carr,~S.; Chen,~S.; Zhang,~C.; et~al., Moiré
  metrology of energy landscapes in van der Waals heterostructures.
  \emph{Nature Communications} \textbf{2021}, \emph{12}, 242\relax
\mciteBstWouldAddEndPuncttrue
\mciteSetBstMidEndSepPunct{\mcitedefaultmidpunct}
{\mcitedefaultendpunct}{\mcitedefaultseppunct}\relax
\EndOfBibitem
\bibitem[Kim \latin{et~al.}(2015)Kim, Kwon, Nikitin, Ahn, Martín-Moreno,
  García-Vidal, Ryu, Min, and Kim]{kim_stacking_2015}
Kim,~D.-S.; Kwon,~H.; Nikitin,~A.~Y.; Ahn,~S.; Martín-Moreno,~L.;
  García-Vidal,~F.~J.; Ryu,~S.; Min,~H.; Kim,~Z.~H. Stacking {Structures} of
  {Few}-{Layer} {Graphene} {Revealed} by {Phase}-{Sensitive} {Infrared}
  {Nanoscopy}. \emph{ACS Nano} \textbf{2015}, \emph{9}, 6765--6773\relax
\mciteBstWouldAddEndPuncttrue
\mciteSetBstMidEndSepPunct{\mcitedefaultmidpunct}
{\mcitedefaultendpunct}{\mcitedefaultseppunct}\relax
\EndOfBibitem
\bibitem[Jeong \latin{et~al.}(2017)Jeong, Choi, Kim, Ahn, Park, Kang, Min,
  Hong, and Kim]{jeong_mapping_2017}
Jeong,~G.; Choi,~B.; Kim,~D.-S.; Ahn,~S.; Park,~B.; Kang,~J.~H.; Min,~H.;
  Hong,~B.~H.; Kim,~Z.~H. Mapping of {Bernal} and non-{Bernal} stacking domains
  in bilayer graphene using infrared nanoscopy. \emph{Nanoscale} \textbf{2017},
  \emph{9}, 4191--4195\relax
\mciteBstWouldAddEndPuncttrue
\mciteSetBstMidEndSepPunct{\mcitedefaultmidpunct}
{\mcitedefaultendpunct}{\mcitedefaultseppunct}\relax
\EndOfBibitem
\bibitem[Wirth \latin{et~al.}(2021)Wirth, Linnenbank, Steinle, Banszerus,
  Icking, Stampfer, Giessen, and Taubner]{wirth_tunable_2021}
Wirth,~K.~G.; Linnenbank,~H.; Steinle,~T.; Banszerus,~L.; Icking,~E.;
  Stampfer,~C.; Giessen,~H.; Taubner,~T. Tunable s-{SNOM} for {Nanoscale}
  {Infrared} {Optical} {Measurement} of {Electronic} {Properties} of {Bilayer}
  {Graphene}. \emph{ACS Photonics} \textbf{2021}, \emph{8}, 418--423\relax
\mciteBstWouldAddEndPuncttrue
\mciteSetBstMidEndSepPunct{\mcitedefaultmidpunct}
{\mcitedefaultendpunct}{\mcitedefaultseppunct}\relax
\EndOfBibitem
\bibitem[Keilmann and Hillenbrand(2004)Keilmann, and
  Hillenbrand]{richards_near-field_2004}
Keilmann,~F.; Hillenbrand,~R. Near-field microscopy by elastic light scattering
  from a tip. \emph{Philosophical Transactions of the Royal Society of London.
  Series A: Mathematical, Physical and Engineering Sciences} \textbf{2004},
  \emph{362}, 787--805\relax
\mciteBstWouldAddEndPuncttrue
\mciteSetBstMidEndSepPunct{\mcitedefaultmidpunct}
{\mcitedefaultendpunct}{\mcitedefaultseppunct}\relax
\EndOfBibitem
\bibitem[Toporski \latin{et~al.}(2018)Toporski, Dieing, and
  Hollricher]{toporski_confocal_2018}
Toporski,~J., Dieing,~T., Hollricher,~O., Eds. \emph{Confocal {Raman}
  {Microscopy}}; Springer {Series} in {Surface} {Sciences}; Springer
  International Publishing: Cham, 2018; Vol.~66\relax
\mciteBstWouldAddEndPuncttrue
\mciteSetBstMidEndSepPunct{\mcitedefaultmidpunct}
{\mcitedefaultendpunct}{\mcitedefaultseppunct}\relax
\EndOfBibitem
\bibitem[Neumann \latin{et~al.}(2015)Neumann, Reichardt, Venezuela,
  Dr{\ifmmode\ddot{o}\else\"{o}\fi}geler, Banszerus, Schmitz, Watanabe,
  Taniguchi, Mauri, Beschoten, Rotkin, and Stampfer]{Neumann2015Sep}
Neumann,~C.; Reichardt,~S.; Venezuela,~P.;
  Dr{\ifmmode\ddot{o}\else\"{o}\fi}geler,~M.; Banszerus,~L.; Schmitz,~M.;
  Watanabe,~K.; Taniguchi,~T.; Mauri,~F.; Beschoten,~B.; Rotkin,~S.~V.;
  Stampfer,~C. {Raman spectroscopy as probe of nanometre-scale strain
  variations in graphene}. \emph{Nat. Commun.} \textbf{2015}, \emph{6},
  8429\relax
\mciteBstWouldAddEndPuncttrue
\mciteSetBstMidEndSepPunct{\mcitedefaultmidpunct}
{\mcitedefaultendpunct}{\mcitedefaultseppunct}\relax
\EndOfBibitem
\bibitem[Malard \latin{et~al.}(2009)Malard, Pimenta, Dresselhaus, and
  Dresselhaus]{Malard_2009}
Malard,~L.; Pimenta,~M.; Dresselhaus,~G.; Dresselhaus,~M. Raman spectroscopy in
  graphene. \emph{Physics Reports} \textbf{2009}, \emph{473}, 51--87\relax
\mciteBstWouldAddEndPuncttrue
\mciteSetBstMidEndSepPunct{\mcitedefaultmidpunct}
{\mcitedefaultendpunct}{\mcitedefaultseppunct}\relax
\EndOfBibitem
\bibitem[Cong \latin{et~al.}(2011)Cong, Yu, Saito, Dresselhaus, and
  Dresselhaus]{Cong_2011}
Cong,~C.; Yu,~T.; Saito,~R.; Dresselhaus,~G.~F.; Dresselhaus,~M.~S.
  {Second-order overtone and combination raman modes of graphene layers in the
  range of 1690-2150 cm-1}. \emph{ACS Nano} \textbf{2011}, \emph{5},
  1600--1605\relax
\mciteBstWouldAddEndPuncttrue
\mciteSetBstMidEndSepPunct{\mcitedefaultmidpunct}
{\mcitedefaultendpunct}{\mcitedefaultseppunct}\relax
\EndOfBibitem
\bibitem[Torche \latin{et~al.}(2017)Torche, Mauri, Charlier, and
  Calandra]{torche_first-principles_2017}
Torche,~A.; Mauri,~F.; Charlier,~J.-C.; Calandra,~M. First-principles
  determination of the {Raman} fingerprint of rhombohedral graphite.
  \emph{Physical Review Materials} \textbf{2017}, \emph{1}, 041001\relax
\mciteBstWouldAddEndPuncttrue
\mciteSetBstMidEndSepPunct{\mcitedefaultmidpunct}
{\mcitedefaultendpunct}{\mcitedefaultseppunct}\relax
\EndOfBibitem
\bibitem[Nguyen \latin{et~al.}(2014)Nguyen, Lee, Yoon, and
  Cheong]{nguyen_excitation_2014}
Nguyen,~T.~A.; Lee,~J.-U.; Yoon,~D.; Cheong,~H. Excitation energy dependent
  {Raman} signatures of {ABA}- and {ABC}-stacked few-layer graphene.
  \emph{Scientific Reports} \textbf{2014}, \emph{4}, 4630\relax
\mciteBstWouldAddEndPuncttrue
\mciteSetBstMidEndSepPunct{\mcitedefaultmidpunct}
{\mcitedefaultendpunct}{\mcitedefaultseppunct}\relax
\EndOfBibitem
\bibitem[Geisenhof \latin{et~al.}(2019)Geisenhof, Winterer, Wakolbinger, Gokus,
  Durmaz, Priesack, Lenz, Keilmann, Watanabe, Taniguchi, Guerrero-Avilés,
  Pelc, Ayuela, and Weitz]{geisenhof_anisotropic_2019}
Geisenhof,~F.~R.; Winterer,~F.; Wakolbinger,~S.; Gokus,~T.~D.; Durmaz,~Y.~C.;
  Priesack,~D.; Lenz,~J.; Keilmann,~F.; Watanabe,~K.; Taniguchi,~T.;
  Guerrero-Avilés,~R.; Pelc,~M.; Ayuela,~A.; Weitz,~R.~T. Anisotropic
  {Strain}-{Induced} {Soliton} {Movement} {Changes} {Stacking} {Order} and
  {Band} {Structure} of {Graphene} {Multilayers}: {Implications} for {Charge}
  {Transport}. \emph{ACS Applied Nano Materials} \textbf{2019}, \emph{2},
  6067--6075\relax
\mciteBstWouldAddEndPuncttrue
\mciteSetBstMidEndSepPunct{\mcitedefaultmidpunct}
{\mcitedefaultendpunct}{\mcitedefaultseppunct}\relax
\EndOfBibitem
\bibitem[Hauer \latin{et~al.}(2012)Hauer, Engelhardt, and
  Taubner]{hauer_quasi-analytical_2012}
Hauer,~B.; Engelhardt,~A.~P.; Taubner,~T. Quasi-analytical model for scattering
  infrared near-field microscopy on layered systems. \emph{Optics Express}
  \textbf{2012}, \emph{20}, 13173--13188\relax
\mciteBstWouldAddEndPuncttrue
\mciteSetBstMidEndSepPunct{\mcitedefaultmidpunct}
{\mcitedefaultendpunct}{\mcitedefaultseppunct}\relax
\EndOfBibitem
\bibitem[Yang \latin{et~al.}(2019)Yang, Zou, Woods, Shi, Yin, Xu, Ozdemir,
  Taniguchi, Watanabe, Geim, Novoselov, Haigh, and Mishchenko]{Yang2019}
Yang,~Y.; Zou,~Y.~C.; Woods,~C.~R.; Shi,~Y.; Yin,~J.; Xu,~S.; Ozdemir,~S.;
  Taniguchi,~T.; Watanabe,~K.; Geim,~A.~K.; Novoselov,~K.~S.; Haigh,~S.~J.;
  Mishchenko,~A. {Stacking Order in Graphite Films Controlled by van der Waals
  Technology}. \emph{Nano Letters} \textbf{2019}, \emph{19}, 8526--8532\relax
\mciteBstWouldAddEndPuncttrue
\mciteSetBstMidEndSepPunct{\mcitedefaultmidpunct}
{\mcitedefaultendpunct}{\mcitedefaultseppunct}\relax
\EndOfBibitem
\bibitem[Nery \latin{et~al.}(2020)Nery, Calandra, and Mauri]{Nery2020}
Nery,~J.~P.; Calandra,~M.; Mauri,~F. {Long-Range Rhombohedral-Stacked Graphene
  through Shear}. \emph{Nano Letters} \textbf{2020}, \emph{20},
  5017--5023\relax
\mciteBstWouldAddEndPuncttrue
\mciteSetBstMidEndSepPunct{\mcitedefaultmidpunct}
{\mcitedefaultendpunct}{\mcitedefaultseppunct}\relax
\EndOfBibitem
\bibitem[Klebl \latin{et~al.}(2022)Klebl, Fischer, Hauck, Wirth, Rothstein,
  Beschoten, Stampfer, Waldecker, Taubner, and Kennes]{InPrep}
Klebl,~L.; Fischer,~A.; Hauck,~J.~B.; Wirth,~K.~G.; Rothstein,~A.;
  Beschoten,~B.; Stampfer,~C.; Waldecker,~L.; Taubner,~T.; Kennes,~D.
  Investigation of possible correlated phases in quadlayer graphene. \emph{In
  preparation} \textbf{2022}, \relax
\mciteBstWouldAddEndPunctfalse
\mciteSetBstMidEndSepPunct{\mcitedefaultmidpunct}
{}{\mcitedefaultseppunct}\relax
\EndOfBibitem
\bibitem[Lu \latin{et~al.}(2021)Lu, Wirth, Gao, Heßler, Sain, Taubner, and
  Zentgraf]{lu_observing_nodate}
Lu,~J.; Wirth,~K.~G.; Gao,~W.; Heßler,~A.; Sain,~B.; Taubner,~T.; Zentgraf,~T.
  Observing {0D} subwavelength-localized modes at {\textasciitilde}100 {THz}
  protected by weak topology. \emph{Science Advances} \textbf{2021}, \emph{7},
  eabl3903\relax
\mciteBstWouldAddEndPuncttrue
\mciteSetBstMidEndSepPunct{\mcitedefaultmidpunct}
{\mcitedefaultendpunct}{\mcitedefaultseppunct}\relax
\EndOfBibitem
\bibitem[Cvitkovic \latin{et~al.}(2007)Cvitkovic, Ocelic, and
  Hillenbrand]{cvitkovic_analytical_2007}
Cvitkovic,~A.; Ocelic,~N.; Hillenbrand,~R. Analytical model for quantitative
  prediction of material contrasts in scattering-type near-field optical
  microscopy. \emph{Optics Express} \textbf{2007}, \emph{15}, 8550\relax
\mciteBstWouldAddEndPuncttrue
\mciteSetBstMidEndSepPunct{\mcitedefaultmidpunct}
{\mcitedefaultendpunct}{\mcitedefaultseppunct}\relax
\EndOfBibitem
\bibitem[Zhan \latin{et~al.}(2013)Zhan, Shi, Dai, Liu, and
  Zi]{zhan_transfer_2013}
Zhan,~T.; Shi,~X.; Dai,~Y.; Liu,~X.; Zi,~J. Transfer matrix method for optics
  in graphene layers. \emph{Journal of Physics: Condensed Matter}
  \textbf{2013}, \emph{25}, 215301\relax
\mciteBstWouldAddEndPuncttrue
\mciteSetBstMidEndSepPunct{\mcitedefaultmidpunct}
{\mcitedefaultendpunct}{\mcitedefaultseppunct}\relax
\EndOfBibitem
\bibitem[Kubo(1957)]{kubo_statistical-mechanical_1957}
Kubo,~R. Statistical-{Mechanical} {Theory} of {Irreversible} {Processes}. {I}.
  {General} {Theory} and {Simple} {Applications} to {Magnetic} and {Conduction}
  {Problems}. \emph{Journal of the Physical Society of Japan} \textbf{1957},
  \emph{12}, 570--586, \_eprint: https://doi.org/10.1143/JPSJ.12.570\relax
\mciteBstWouldAddEndPuncttrue
\mciteSetBstMidEndSepPunct{\mcitedefaultmidpunct}
{\mcitedefaultendpunct}{\mcitedefaultseppunct}\relax
\EndOfBibitem
\bibitem[Gr\"uneis \latin{et~al.}(2008)Gr\"uneis, Attaccalite, Wirtz, Shiozawa,
  Saito, Pichler, and Rubio]{Rubio_flg_Tb}
Gr\"uneis,~A.; Attaccalite,~C.; Wirtz,~L.; Shiozawa,~H.; Saito,~R.;
  Pichler,~T.; Rubio,~A. Tight-binding description of the quasiparticle
  dispersion of graphite and few-layer graphene. \emph{Phys. Rev. B}
  \textbf{2008}, \emph{78}, 205425\relax
\mciteBstWouldAddEndPuncttrue
\mciteSetBstMidEndSepPunct{\mcitedefaultmidpunct}
{\mcitedefaultendpunct}{\mcitedefaultseppunct}\relax
\EndOfBibitem
\end{mcitethebibliography}

\newpage

\begin{center}
\textbf{\large Supplementary material: Experimental observation of ABCB stacked tetralayer graphene
}
\end{center}

\setcounter{equation}{0}
\setcounter{figure}{0}
\setcounter{table}{0}
\setcounter{page}{1}
\makeatletter
\renewcommand{\theequation}{S\arabic{equation}}
\renewcommand{\thefigure}{S\arabic{figure}}

\subsection*{S1 Stability and abundance}
\begin{figure}[!bht]
    \begin{center}
        \includegraphics[width=0.45\textwidth]{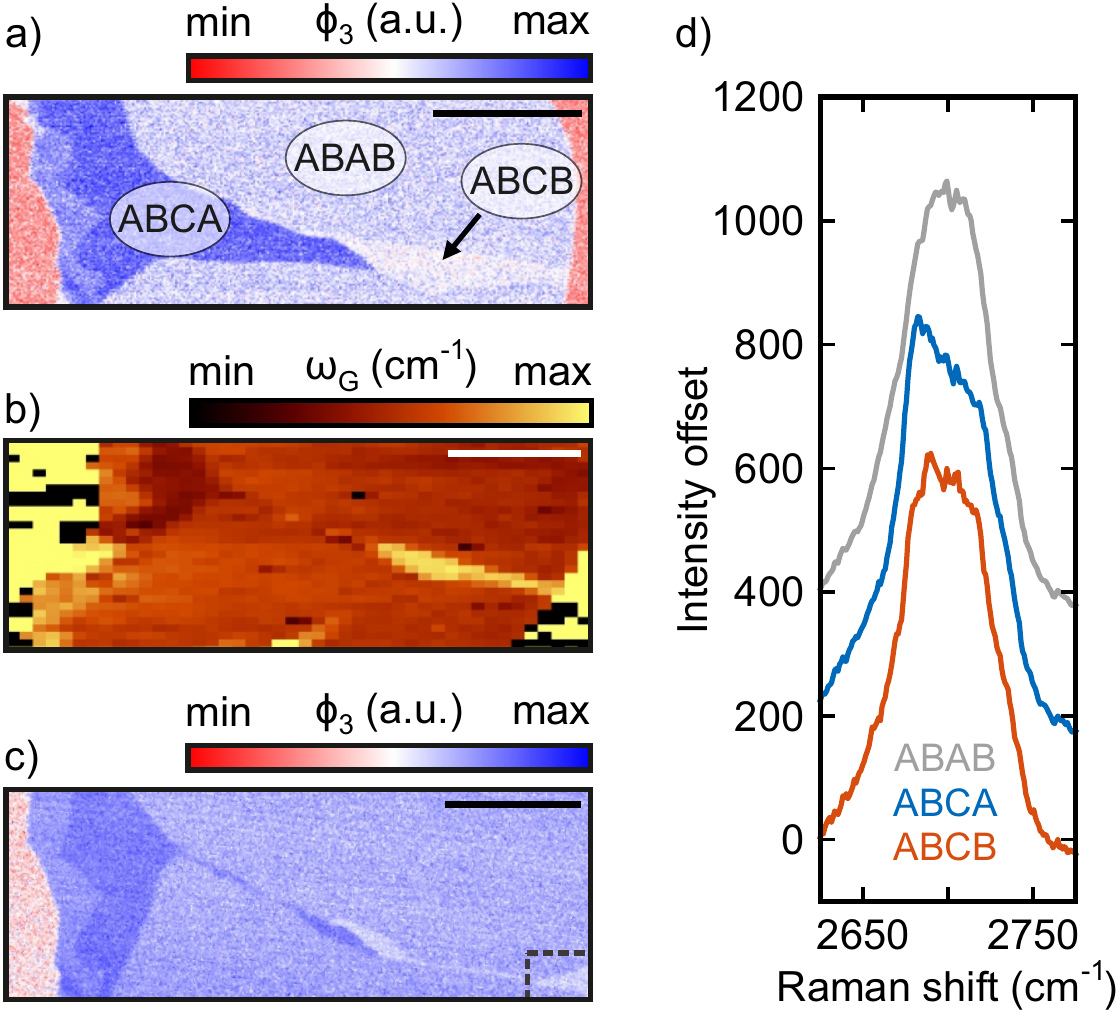}
        \caption{ Flake 2: a) Optical phase ($\Phi_3$) image of a 4LG flake on SiO\textsubscript{2}, recorded first at 0.34~{eV}. b) Raman map around the 2D peak, recorded after the s-SNOM measurement with a high intensity Raman laser reveals a shrinkage of ABCA domain.  c) Subsequent optical phase ($\Phi_3$) image obtained at photon energies of 0.34~{eV}, after the Raman measurements, reveals a shrinkage of the ABCB domain when compared to  b) and a). The dashed box indicates the region where s-SNOM nano-spectroscopy was conducted. d) Raman spectra recorded in the different regions for the three domains in b). The scalebars correspond to 5 \textmu m each.}
        \label{fig:S1_2}
    \end{center}
\end{figure}

In Figure ~\ref{fig:S1_2} the optical phase ($\Phi_3$) and a Raman map of the 4LG flake labeled Flake 2 in the main text are shown over the course of several measurements.
The phase image ($\Phi_3$) in Figure ~\ref{fig:S1_2} a) is recorded at a photon energy of 0.34~{eV}. The 4LG flake can be clearly distinguished from the SiO\textsubscript{2} substrate on the left (red area). Three regions on the 4LG are marked with their respective stacking. 
The domain with the weakest contrast towards the SiO\textsubscript{2} is ABCB and the strongest ABCA.

The Raman map around the 2D peak recorded after the s-SNOM measurements is depicted in Figure~\ref{fig:S1_2} b). 
The ABCA-domain shrunk in the middle of the image, while the ABCB-domain is still present. 
However, due to the limited resolution of the Raman setup, the exact shape of the domains is difficult to determine. 
Subsequent s-SNOM measurements repeated at 0.34~{eV} are depicted in Figure ~\ref{fig:S1_2} c). 
The ABCA-domain has a similar size to what can be expected from the Raman in Figure ~\ref{fig:S1_2} b), while the ABCB-domain shrunk again to a small stripe and a triangluar feature at the right side of the image. 
This feature is indicated by a black dashed rectangle, which corresponds to the region investigated with IR nano-spectroscopy.
In Figure~\ref{fig:S1_2} d) Raman spectra around the 2D peak corresponding to the map in Figure ~\ref{fig:S1_2} b) are shown. 
The different domains exhibit similar features like in Figure~3 b).

\begin{figure}[!bht]
    \begin{center}
        \includegraphics[width=0.45\textwidth]{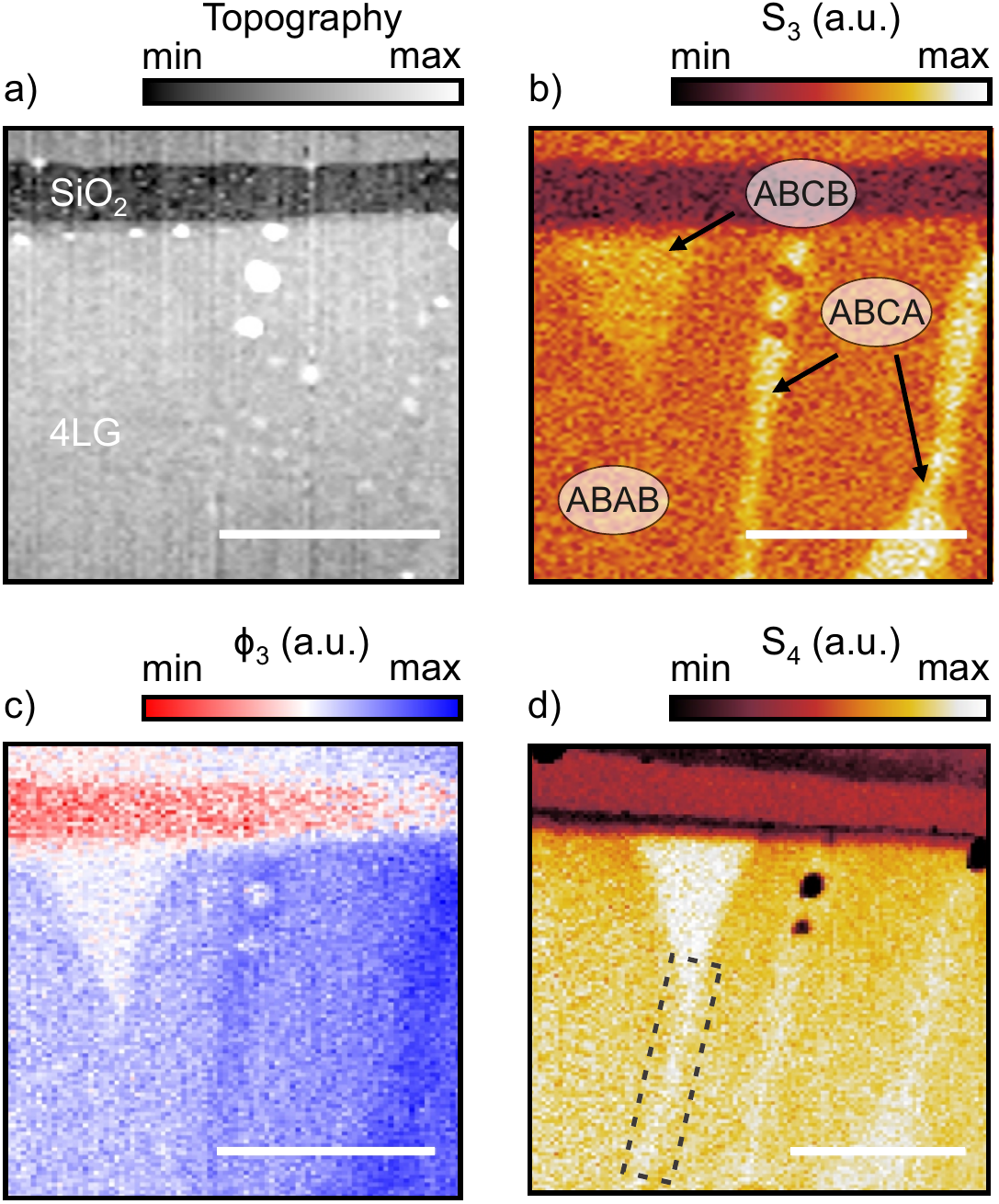}
        \caption{ Flake 3: a) Topography image of the 4LG-only flake on SiO\textsubscript{2}. b), c) Corresponding optical amplitude (S\textsubscript{3}) and phase ($\Phi_3$) images obtained at photon energies of 0.34~{eV}, revealing regions of different s-SNOM signal, with three different domains indicated by ABAB, ABCA and ABCB. The domains correspond to the same domains as in Figure~2  of the main text. d) Amplitude image obtained at 10.6 µm of the same area. A bright line below the ABCB domain is visible. 
        The scale bar corresponds to 1 µm.}
        \label{fig:S1_1}
    \end{center}
\end{figure}

In Figure ~\ref{fig:S1_1} a) the topography of the 4LG flake labeled as Flake 3 in the main text is shown. 
The topography image of the 4LG is featureless, only a few dirt particles and the SiO\textsubscript{2} substrate can be identified, indicating a homogenous number of four graphene layers.

In the corresponding optical amplitude (S\textsubscript{3}) and phase image ($\Phi_3$) in Figure~\ref{fig:S1_1} b) and c) recorded at a photon energy of 0.34~{eV} the flake can be clearly distinguished from the SiO\textsubscript{2} substrate. 
It shows three different amplitude and phase signals on the topographically featureless 4LG. The corresponding domains are indicated by ABAB, ABCA and ABCB. ABAB-domains cover the largest part of the 4LG . 
Two small domains of the largest contrast correspond to ABCA stacking. 
The triangular shaped ABCB-domain is located at the edge between 4LG and the SiO\textsubscript{2} and has an amplitude signal between those of ABAB- and ABCA-domain. 
In the phase image in Figure~\ref{fig:S1_1}  c) the same three domains can be identified. 
Again, the phase response is the largest with respect to the SiO\textsubscript{2} for the ABCA-domain and weakest for ABCB.

During our investigations, the domains of Flake 1 (main text) remained stable over several s-SNOM and Raman measurements (over the course of several weeks). 
For Flake 2 shown in Figure~\ref{fig:S1_2}, however, we observed the consecutive collapse of a large ABCB domain to a small triangular shaped domains after Raman mapping at laser powers in the range of 3 mW. 
Such a change in the stacking order has been reported for TLG \cite{zhang_light-induced_2020}, albeit for significantly higher laser power. 
The third investigated flake (Flake 3, Figure~\ref{fig:S1_1}) also shows a triangular shaped similar sized domain of ABCB stacking, possibly a remnant of a collapsed larger domain. 
This argument is supported by investigations of the respective area with s-SNOM at a wavelength of 10.6~{µm} (Figure~\ref{fig:S1_1} d)), which reveals a small boundary like feature, indicated by the black box, similar to a shear soliton observed in bilayer graphene ~\cite{jiang_soliton-dependent_2016} and TLG~\cite{jiang_manipulation_2018}.
The energy barrier between ABCB and ABAB stacking is expected to be much lower compared to the transition from ABCA to ABAB~\cite{aoki_dependence_2007}. This instability might hamper device fabrication, because the ABCB stacking can transform to energetically favorable Bernal stacking, similar to metastable rhombohedral graphene upon stress or strain during device fabrication~\cite{geisenhof_anisotropic_2019}.

\subsection{S2 s-SNOM contrast calculation}
The s-SNOM contrast for the 4LG/SiO$_2$/Si layer stack is calculated with the finite dipole model~\cite{cvitkovic_analytical_2007} with an extension for layered samples~\cite{hauer_quasi-analytical_2012} as described in \cite{wirth_tunable_2021}.

The effective polarizability of the tip-sample system is calculated via
\begin{equation}
    \alpha_{\mathrm{eff}} = W_0^2L \frac{\frac{2L}{a}+\ln{\frac{a}{4e L}}}{\ln{\frac{4L}{e^2a}}}(2+\eta_r(z)),
\end{equation}

where $a$ is the tip radius, $L$ is the effective length of the assumed sphere and $W_0=1.31a$. The detailed description of the model can be found in \cite{hauer_quasi-analytical_2012, cvitkovic_analytical_2007}. 
\begin{equation}
    \eta_r(z) = \frac{\beta (2Lg-2H-W_0-a)\ln(\frac{4L}{4H+2W_0+a})}{4L \ln(\frac{4L}{a})-\beta(4Lg-4H-3a)\ln(\frac{4L}{4H+2a})},
\end{equation}
where $H$ is the tip height above the sample and g corresponds to a fraction of the total induced charge. The n-th Fourier component needs to be included to account for the higher harmonic demodulation of $\alpha_{\mathrm{eff}}$. The final amplitude and phase contrast depend solely on the third order Fourier component of  $\eta_r(z)$ because the far-field coefficients can be neglected here.  
\begin{equation}
    \frac{S_3}{S_{3}^{\textrm{ref}}} = \frac{\textrm{Abs}[(\eta_3)]}{\textrm{Abs}[(\eta_3^{\textrm{ref}})]}
\end{equation}

\begin{equation}
    \Phi_3-\Phi_3^{\textrm{ref}} = \textrm{Arg}[(\eta_3)]-\textrm{Arg}[(\eta_3^{\textrm{ref}})]
\end{equation}
Parameters of the FDM are summarized in Table~\ref{tab:FDMParams}.
We replace the electrostatic reflection coefficient $\beta$ with a Fresnel reflection coefficient \cite{hauer_quasi-analytical_2012} for a single dominant in-plane wave vector $k_{||}=250000$~cm$^{-1}$, comparable to the inverse of the expected tip radius of $\rho=3.3\cdot10^5$~cm$^{-1}$.
To calculate the Fresnel coefficients for the stack, we use the transfer matrix method (TMM) for graphene layers with p-polarized light \cite{zhan_transfer_2013}. In our model the tetralayer graphene is infinitesimal thin and a plane interface.

\begin{table}[!ht]
    \centering
\caption{FDM Parameter}\label{tab:FDMParams}
\begin{tabular}{|c|c|}
    \hline
    Name & Value \\
    \hline
    Demodulation order $n$ & $3$ \\
    Tapping Amplitude $H$ & $50$~{nm} \\
    $L$~\cite{cvitkovic_analytical_2007} & $300$~{nm}  \\
    $g$~\cite{cvitkovic_analytical_2007} & $0.7e^{(i0.06)}$ \\
    Tip radius $a$ & $30$~{nm}\\
    SiO\textsubscript{2} thickness & $90$~{nm} \\
     \hline
\end{tabular}
\end{table}

\subsection*{S3 s-SNOM contrasts for different chemical potentials}

\begin{figure}[!bht]
    \begin{center}
        \includegraphics[width=0.45\textwidth]{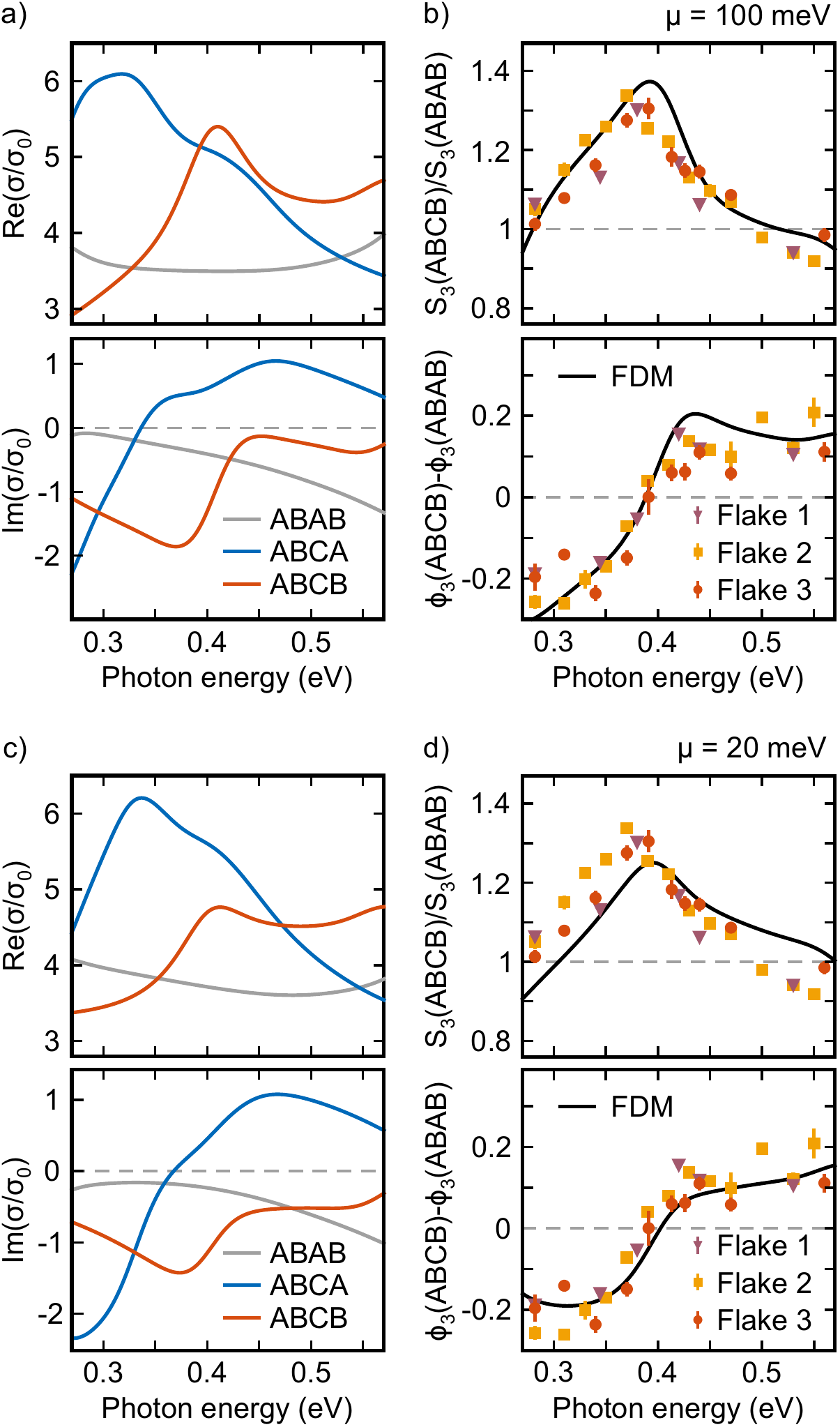}
        \caption{Optical conductivity data for ABCB plotted for $\mu=20$~{meV} a) and $\mu=100$~{meV} c). Comparison between experimental data for the three flakes and the modelled calculation for the chemical potentials.}
        \label{fig:S3}
    \end{center}
\end{figure}

In  Figure ~\ref{fig:S3} a) and c) the optical conductivity of the three stackings is plotted, calculated for a broadening of $\eta=40$~{meV} and chemical potentials $\mu$ of $20$ and $100$~{meV}, respectively. 
In b) and d) same amplitude and phase data as in Figure~4 are plotted for ABCB referenced to ABAB. 
The calculated FDM spectra are shown for two different chemical potential. 
All three amplitude data sets show a higher peak at 0.38~{eV}, which can be either attributed to a smaller broadening or an increased chemical potential.
For a change in the chemical potential, we expect a shift of the characteristic crossing of 1 and 0 in amplitude and phase, respectively. 
Furthermore, this also influences the contrast strength. 
A change in $\eta$ would result in more pronounced features such as the peak in amplitude around 0.4~{eV}. 
For Flake 1 and Flake 2 the data fit better to 100~{meV}, especially the crossing of 1 in the amplitude is better reproduced. 
There are also indications for a smaller broadening, because amplitude data of Flake 1 indicate a sharper left flank and Flake 2 a more pronounced peak.

\subsection*{S4 Calculation of optical conductivity}
For the calculation of the dynamic optical conductivity we employ the Kubo formula~\cite{kubo_statistical-mechanical_1957}
\begin{equation}
    \sigma(\omega)_{i,j} = \frac{e^2\hbar}{iS} \sum_{b_1,b_2,\bvec k} \frac{n_f(e_{b_1}^{\bvec k})-n_f(e_{b_2}^{\bvec k})}{e_{b_1}^{\bvec k}-e_{b_2}^{\bvec k}} \\
    \frac{\Braket{e_{b_1}^{\bvec k}|\mathbf{v}_i|e_{b_2}^{\bvec k}}\Braket{e_{b_2}^{\bvec k}|\mathbf{v}_j|e_{b_1}^{\bvec k}}}{\hbar\omega + e_{b_1}^{\bvec k}-e_{b_2}^{\bvec k} + i\eta} \,,
\end{equation}
where $e_b(\bvec k)$ are the eigenenergies of the Hamiltonian in band $b$ at momentum point $\bvec k$ and $\ket{e_{b}^{\bvec k}}$ are the corresponding eigenvectors. The case of $e_{b_1}^{\bvec k} = e_{b_2}^{\bvec k}$ has to be considered separately and results in $\beta n_f(e_{b_1}^{\bvec k})(n_f(e_{b_1}^{\bvec k})-1)$. $\mathbf{v}_i(\bvec k)$ is the $i$-direction component of the velocity operator defined as $\mathbf{v}^n = \frac{i}{\hbar}[\mathbf{H},\mathbf{x}^n]$ which can be expressed in the momentum-site basis as
\begin{equation}
    \mathbf{v}^n_{o_1,o_2} =  -\frac{i}{\hbar}\sum_{\bvec d}e^{-i \bvec k \bvec d} H_{\bvec{r}_{1},\bvec{r}_{2}+\bvec d}\cdot (r_{o_1}^n-r_{o_2}^n+\bvec d^n),
\end{equation} 
with $\bvec{r_i}$ a vector pointing to a site $i$ in the lattice and $\bvec{d}$ being a lattice vector. We chose the temperature to be room temperature ($0.025$\,eV) and choose the broadening $\eta$ as $40$\,meV. We chose a simple tight-binding Hamiltonian of the form
\begin{equation}
    H = \sum_{i,j}\delta_{|\bvec{r}_i-\bvec{r}_j|,d_{cc}}\gamma_{\mathrm{intra}} c_{i}^{\dagger}c_{j}+ \delta_{|\bvec{r}_i-\bvec{r}_j|,d_{\mathrm{layer}}}\gamma_{\mathrm{inter}} c_{i}^{\dagger}c_{j} 
\end{equation}
consisting of only nearest-neighbor intralayer (in a distance $d_{cc} = 1.42$\,\AA) and nearest-neighbor interlayer hopping (in a distance $d_{\mathrm{layer}} = 3.35$\,\AA) with hopping energies of $\gamma_{\mathrm{intra}} = 3.16$\,eV and $\gamma_{\mathrm{inter}} = 0.39$\,eV. We shift the Fermi energy to $50$\,meV. The modeling parameters, Fermi-energy and broadening $\eta$ were tweaked for the best agreement with the experimental results. 

For the calculations of the DOS and the bandstructure in Fig.~\ref{fig:0}, we employ a Slonzecewski-Weiss-McClure (SWMC) Hamiltonian, with parameters chosen based on Ref.~\cite{Rubio_flg_Tb} summarized in Table~\ref{tab:TBparams}. The Hamilton-matrix can be constructed as
\begin{equation}
    H_{i,j}^{y}(\bvec k) = \delta_{i,j}\delta_{i,A} \Delta_{y} \\
    +\sum_{u_1,u_2 \in \mathbb{Z}} \sum_n \,\delta_{d_{i,j}^{u_1,u_2},d_n}e^{-i\bvec k \bvec R(u_1,u_2)}\gamma_n \,,
\end{equation}
where $i$ and $j$ are site indices within the unit cell, $y$ marks the type of the lattice and is either ABAB, ABCA or ABCB and $\delta_{i,A}$ is one if $i$ is an $A$-site and $0$ otherwise. $u_1$ and $u_2$ iterate over all unit cells in the infinite lattice, $\bvec R(u_1,u_2)$ gives the vectorial distance between the unit cells and $d_{i,j}^{u_1,u_2}$ gives the distance between site $i$ and the image of site $j$ shifted by $u_1$ and $u_2$ unit-cell vectors. $d_n$ is the real-space distance associated with each of the hopping parameters is listed in the third column of Table~\ref{tab:TBparams}.
\begin{table}[!tbh]
    \centering
\caption{SWMC model parameters}\label{tab:TBparams}
\begin{tabular}{|c|c|c|}
    \hline
    Name & Value in eV & Distance in \AA \\
    \hline
    $\gamma_0$ & $2.553$ & $1.42$\\
    $\gamma_1$ & $0.343$ & $3.35$\\
    $\gamma_2$ & $-0.009$ & $6.70$ \\
    $\gamma_3$ & $0.18$ &  $4.16$\\
    $\gamma_4$ & $0.173$ & $3.64$\\
    $\gamma_5$ & $0.018$ & $6.85$\\
    $\Delta_{\text{ABAB}}$ &  $-0.003$ & $0.0$, A site\\ 
    $\Delta_{\text{ABCA}} $&  $0.0$& $0.0$, A site\\
    $\Delta_{\text{ABCB}} $& $-0.018$ & $0.0$, A site\\
     \hline
\end{tabular}
\end{table}

The values for the $\Delta_y$ are chosen such that we roughly reproduce the low energy behavior of the bandstructures from Ref.~\cite{aoki_dependence_2007}. Therefore, the model is a good representation of the low energy degrees of freedom near half-filling. 

\subsection*{S5 FTIR measurements of tetralayer graphene}
\begin{figure}[!bht]
    \begin{center}
        \includegraphics[width=1\textwidth]{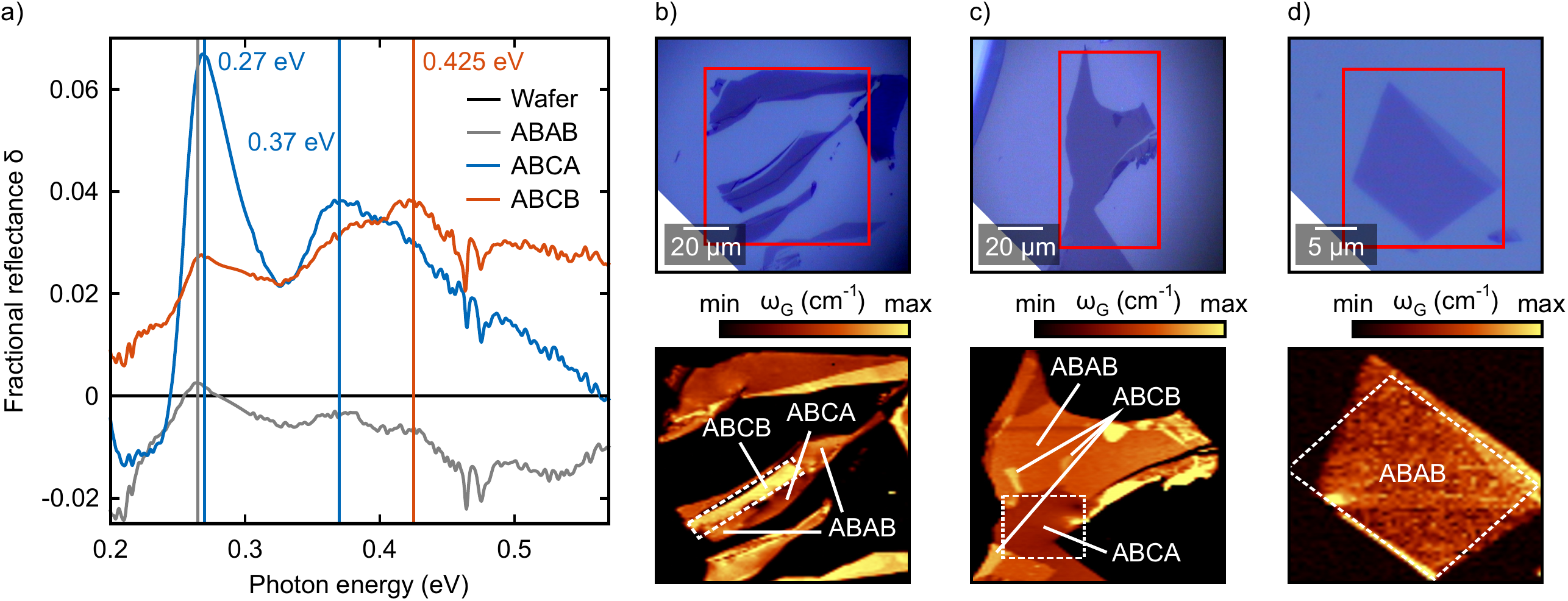}
        \caption{a) Fractional change of reflectance for three different stacking orders measured by FTIR with small apertures. The measurements were conducted under ambient condition. The CO\textsubscript{2} absorption peak between 0.27 and 0.3~{eV}  was removed by interpolation. b), c) and d) Optical microscopy (top row) and Raman images (bottom row) of the investigated flakes. The red boxes in the optical images indicate the position of the Raman images. The white boxes in b), c) and d) indicate the areas (aperture position) where the FTIR spectra in a) were recorded. Different flakes were chosen to maximize the signal from the respective areas. Note that the flakes shown here are different from the ones shown in previous figures.}
        \label{fig:SFTIR}
    \end{center}
\end{figure}

The optical properties of few layer graphene are determined by its optical conductivity, which differs between the different crystal polytypes\cite{mak_electronic_2010}.
Therefore, the stacking order in few layer graphene can be determined by infrared far-field spectroscopy \cite{mak_electronic_2010, mak_evolution_2010} in the range between 0.2 and 1~{eV}. 
The influence of environmental effects, such as doping, on the infrared response of tetralayer graphene is expected to be reduced compared to thinner flakes \cite{mak_electronic_2010}, making far-field spectroscopy a reliable technique.
Requirement for the investigation with this method, however, are domains of sufficient size due to the optical diffraction limit.
Within a set of approximately 50 flakes scanned by Raman spectroscopy, we identified one suitable ABCB domain.
The flake (shown in Fig \ref{fig:SFTIR} b) exhibits an ABCB domains of approximately $5 \cdot 30$~\textmu m$^2$, adjacent to both ABCA and ABAB stacking.

Fourier transform infrared spectroscopy (FTIR) measurements on the three flakes shown in \ref{fig:SFTIR} reveal the infrared response of ABAB, ABCA and ABCB stacking orders.
All measurements were performed in reflection geometry as the flakes are exfoliated onto SiO\textsubscript{2}/Si substrates. 
We calculate the fractional change of the reflectance as following:

\begin{equation}
    \Delta R=\dfrac{R_{4LG}-R_s}{R_s},
\end{equation}

where $R_{4LG}$ is the reflectance of the flake and $R_s$ is the reflectance of the substrate.
The ABAB stacking is featureless and exhibits only two small peaks at 0.26~{eV} and above 0.6~{eV}.
The ABCA stacking exhibits two peaks, a pronounced one at 0.27~{eV} and a weaker one at 0.37~{eV}.
This in good agreement with literature data on ABAB and ABCA stacked 4LG \cite{mak_electronic_2010}.
The fractional reflectance of ABCB stacking has a peak at 0.425~{eV}, which can be distinguished from the higher energy peak of ABCA by a shift towards higher energy and a slightly higher amplitude. The prominent peak of ABCA at lower energies is absent.
The peak at 0.425~{eV} agrees well with the peak position in the real part of the conductivity (c.f. Figure \ref{fig:2}a).
The same peak as in the ABAB stacking at 0.26~{eV} is also present, because the surrounding of the ABCB stacking consists mostly of ABAB stacking which also contributes to the signal due to the non perfect aperture and the diffraction limit of infrared radiation.
The peaks in the FTIR spectra correspond to the splitting between the conduction bands and are thus unique for each stacking order \cite{mak_electronic_2010, mak_evolution_2010}. This is the same principle as for the s-SNOM measurements presented in Figure \ref{fig:2} of the main text, but lacking the sub-diffraction limit spatial resolution of s-SNOM.

\end{document}